\documentclass[twocolumn]{aastex62}

\newcommand{\gadget}{\textsc{Gadget-3}}
\newcommand{\sphgal}{\textsc{SPHGal}}
\newcommand{\subfind}{\textsc{Subfind}}
\newcommand{\treecol}{\textsc{TreeCol}}
\newcommand{\healpix}{\textsc{healpix}}
\newcommand{\basel}{\textsc{BaSeL}}

\graphicspath{{./}{figures/}}

\received{12 Nov 2019}
\revised{17 Jan 2020}
\accepted{28 Jan 2020}
\submitjournal{ApJ}

\shorttitle{}
\shortauthors{Lah\'{e}n et al.}

\begin{document}

\title{ The \textsc{griffin} project - Formation of star clusters with individual massive stars in a simulated dwarf galaxy starburst}

\correspondingauthor{Natalia Lah\'{e}n}
\email{natalia.lahen@helsinki.fi}

\author[0000-0003-2166-1935]{Natalia Lah\'{e}n}
\affil{Department of Physics, University of Helsinki, Gustaf H\"allstr\"omin katu 2, FI-00014 Helsinki, Finland}

\author{Thorsten Naab}
\affiliation{Max-Planck-Institut f\"ur Astrophysik, Karl-Schwarzschild-Str. 1, D-85748, Garching, Germany}

\author{Peter H. Johansson}
\affiliation{Department of Physics, University of Helsinki, Gustaf H\"allstr\"omin katu 2, FI-00014 Helsinki, Finland}

\author{Bruce Elmegreen}
\affiliation{IBM T. J. Watson Research Center, 1101 Kitchawan Road, Yorktown Heights, NY 10598, USA}

\author{Chia-Yu Hu}
\affiliation{Max-Planck-Institut f\"ur extraterrestrische Physik, Giessenbachstr. 1, D-85748, Garching, Germany}
\affiliation{Center for Computational Astrophysics, Flatiron Institute, 162 5th Ave, New York, NY 10010, USA}
\author{Stefanie Walch}
\affiliation{I. Physikalisches Institut, Universit\"at zu K\"oln, Z\"ulpicher Strasse 77, D-50937 K\"oln, Germany}

\author{Ulrich P. Steinwandel}
\affiliation{Universit\"ats-Sternwarte M\"unchen, Fakult\"at f\"ur Physik, LMU M\"unchen, Scheinerstr. 1, 81679 Munich, Germany}
\affiliation{Max Planck Institute for Astrophysics, Karl-Schwarzschild-Str. 1, D-85740, Garching, Germany}

\author{Benjamin P. Moster}
\affiliation{Universit\"ats-Sternwarte M\"unchen, Fakult\"at f\"ur Physik, LMU M\"unchen, Scheinerstr. 1, 81679 Munich, Germany}
\affiliation{Max Planck Institute for Astrophysics, Karl-Schwarzschild-Str. 1, D-85740, Garching, Germany}

\begin{abstract}

We describe a population of young star clusters (SCs) formed in a hydrodynamical simulation of a gas-rich dwarf galaxy merger resolved with individual massive stars at sub-parsec spatial resolution. The simulation is part of the \textsc{griffin} (Galaxy Realizations Including Feedback From INdividual massive stars) project. 
The star formation environment during the simulation spans seven orders of magnitude in gas surface
density and thermal pressure, and the global star formation rate surface density ($\Sigma_\mathrm{SFR}$) varies by more than three orders of magnitude during the simulation. 
Young SCs more massive than $M_{\mathrm{*,cl}}\sim 10^{2.5}\,M_{\odot}$ form along a mass function with a power-law index $\alpha\sim-1.7$ ($\alpha\sim-2$ for $M_{\mathrm{*,cl}}\gtrsim10^{3}\,M_{\odot}$) at all merger phases, while the normalization and the highest SC masses (up to $\sim 10^6 M_{\odot}$) correlate with $\Sigma_\mathrm{SFR}$. The cluster formation efficiency varies from $\Gamma\sim20\%$ in early merger phases to $\Gamma\sim80\%$ at the peak of the starburst and is compared to observations and model predictions. The massive SCs ($\gtrsim10^4\,M_{\odot}$) have sizes and mean surface densities similar to observed young massive SCs. Simulated lower mass clusters appear slightly more concentrated than observed. 
All SCs form on timescales of a few Myr and lose their gas rapidly resulting in typical stellar age spreads between $\sigma\sim0.1-2$ Myr ($1\sigma$), consistent with observations. The age spreads increase with cluster mass, with the most massive cluster ($\sim10^6\, M_{\odot}$) reaching a spread of $5\, \mathrm{Myr}$ once its hierarchical formation finishes. Our study shows that it is now feasible to investigate the SC population of entire galaxies with novel high-resolution numerical simulations.

\end{abstract}


\section{Introduction}

The majority of stars in galaxies form in hierarchical patterns \citep[e.g.][]{2010ARA&A..48..339B,2017ApJ...840..113G} with dense clusters \citep{2003ARA&A..41...57L,2010ARA&A..48..431P} growing from sub-cluster coagulation and gas accretion on the smallest scales \citep{2007ApJ...655L..45M,2008ApJ...672.1006E} and dispersing with increasing age \citep{2013MNRAS.430..676B,2018MNRAS.475.5659W}.

As a result of this hierarchy, star clusters typically have mass functions (CMF) $dN/dM \propto M^\alpha$ with power-law slopes of $\alpha \sim -2$ independent of environment (e.g. \citealt{1996ApJ...471..816E,1996ApJ...466..802E,1999ApJ...527L..81Z,2003AJ....126.1836H,2007MNRAS.380.1271P, 2012ApJ...752...96F}). The lower cluster mass limit of $\sim 10^2 M_{\odot}$ is set by the masses of individual stars, while some authors find evidence for an extended power-law distribution of clusters all the way to very high masses of $10^6 M_{\odot} \lesssim M_* \lesssim 10^7 M_{\odot}$ (e.g. \citealt{2019ApJ...872...93M}). In the case of the extended power-law CMF, globular clusters with typical masses of $M_* \sim 10^5 M_{\odot}$ would just be the massive relics of normal star clusters forming in extreme environments at high redshift (\citealt{1997ApJ...480..235E,2001ApJ...561..751F}; \citealt{2015MNRAS.454.1658K,2019MNRAS.483.3618V}). On the other hand, several studies imply an upper cutoff around $M_* \sim 10^5 M_{\odot}$ \citep{2010ARA&A..48..431P,2015MNRAS.452..246A,2018MNRAS.477.1683M}, making the CMF resemble a Schechter function, which is described by a power-law with an exponential tail.

In the local Universe the formation of star clusters is enhanced in extreme environments, such as galaxy mergers, starburst galaxies and galactic nuclei (e.g. \citealt{1995AJ....110.2665M,1999AJ....118.1551W,2005A&A...431..905B,2010RvMP...82.3121G}; \citealt{2019MNRAS.490.1714P}). It is debated in the observational literature, whether the cluster formation rate follows the global star formation rate, independent of galaxy type \citep[e.g.][]{2015ApJ...810....1C,2017ApJ...849..128C} or if the efficiency for young clusters surviving the embedded phase increases for galaxies with higher star formation rates (e.g. \citealt{2010MNRAS.405..857G,2012MNRAS.426.3008K}; \citealt{2019MNRAS.490.1714P}).  In the first case, the fraction of young stars that are in clusters is independent of the star formation rate per unit area, whereas in the second case, the fraction increases with the star formation rate per unit area.

The formation of the population of young massive clusters (YMC) is of particular interest. They are very compact, with 
half-light radii of only a few pc and  masses of $\sim 10^{4}-10^{8} M_{\odot}$, which is similar to the 
masses of present-day globular clusters (e.g. \citealt{2004A&A...416..537L, 2010ARA&A..48..431P, 2010ApJ...716L..90R, 2012A&A...539A...5C}). Given their many similarities, many authors have suggested that the present-day globular clusters are just YMCs formed at early times \citep{2014prpl.conf..291L,2014CQGra..31x4006K}. GCs would then be the fossil record of an intense 
early episode of clustered star formation, which is evidenced in the present-day population of globular clusters found 
ubiquitously in the haloes of all types of galaxies, even in low-mass dwarf galaxies (e.g. \citealt{1991ARA&A..29..543H,2018ARA&A..56...83B}).

The observations of YMCs demonstrate that they are preferentially formed in very dense and gas-rich environments characterized by 
strongly turbulent velocity fields and very high gas pressures of $P_{\rm th} \gtrsim 10^{7}\, \rm k_{B}\,(K\, cm^{-3})^{-1}$. Such high densities and pressures, uncommon in the local Universe but more prevalent in the high redshift Universe, also 
result in elevated integrated star formation efficiencies, in which a larger fraction of the mass of a given molecular 
cloud might be turned into stars (e.g. \citealt{2010ApJ...710L.142F,2015ApJ...809..187S,2019MNRAS.487..364L}). The increased star formation efficiency in turn would also result in a higher cluster formation efficiency (CFE), which describes the fraction of star formation occurring in bound stellar clusters (e.g. \citealt{1997ApJ...480..235E,2012MNRAS.426.3008K}). Even if the cluster formation efficiency is independent of environment, higher gas surface densities and therefore star formation rate surfaces densities would result in a higher normalization of the CMF and therefore an increase in the peak cluster mass \citep[e.g.][]{2014ApJ...795..156W}. Thus, a mechanism that can generate high gas densities, such as a galaxy merger, will be most beneficial for studying a large population of stellar clusters. 

Traditionally, star cluster evolution has been investigated using direct $N$-body simulations that resolve the internal evolution of pre-existing star clusters set in a tidal field at high spatial and temporal precision \citep{2007ApJ...655L..45M,2011MNRAS.418..759R,2016MNRAS.458.1450W}. However, the direct $N$-body simulations typically do not 
include hydrodynamical processes, which are crucial for studying the actual formation process of stellar clusters as stars form
in dense molecular clouds. In addition, due to the steep scaling of computational time $O(N^{2})$ with particle number $N$, these simulations are often unable to model the environment around the stellar clusters and therefore lack the detailed interplay between the clusters and their host galaxy. 

To date, most of the galactic-scale numerical simulation work has concentrated on simulating the formation of massive clusters ($>10^3\, M_\odot$) such as proto-globular clusters \citep{2015MNRAS.446.2038R,2018MNRAS.474.4232K,2019arXiv190611261M,2018NewAR..81....1R} with the stars represented as population particles resolved down to $10^2 M_{\odot}$ \citep[see e.g.][for recent improvements towards realizations of individual massive stars]{2017MNRAS.471.2151H,2018ApJ...865L..22E}. Observations in the Milky Way probe cluster masses down to a minimum mass of only a few tens of solar masses \citep{2003ARA&A..41...57L,2006AJ....131.1559V}. One of the major challenges in modelling the formation of stellar clusters in a realistic galactic environment are their very compact sizes. The effective radii of clusters are typically in the range of only $\sim 1-10$ parsecs \citep{1994ApJ...433...65O, 2005ApJS..161..304M,2013MNRAS.431.1252B}, thus resolving their formation process will require 
high spatial resolution. 

Numerically the formation of stellar clusters has been studied using isolated simulations of collapsing and fragmenting
molecular clouds (e.g. \citealt{2001ApJ...556..837K,2003MNRAS.343..413B,2017MNRAS.467.3255M}; \citealt{2017ApJ...840...48P}, \citealt{2017MNRAS.467.1313V,2018NatAs...2..725H}). In addition, the properties of star clusters have also been studied in stratified disk models \citep[e.g.][]{2015MNRAS.454..238W,2017MNRAS.466.3293P,2018ApJ...853..173K}, in spiral galaxies and their mergers (e.g. \citealt{2002MNRAS.335.1176B,2004ApJ...614L..29L,2005ApJ...626..823L,2008MNRAS.389L...8B}) and in dwarf galaxies (e.g. \citealt{2010ASPC..423..185S,2013MNRAS.430.1901H}), however typically without resolving their internal structure. Idealised merger simulations provide a high spatial and mass resolution \citep{2015MNRAS.446.2038R,2018MNRAS.475.4252A}, even down to solar masses per resolution element \citep{2019ApJ...879L..18L}, but lack the cosmological environment provided by cosmological simulations.
Meanwhile current cosmological zoom-in simulations have been used to gain insight in the formation of stellar clusters. However, the highest available mass resolution is still of the order of hundreds of solar masses \citep{2019arXiv190611261M}. Interpreting the formation of stellar clusters in a full cosmological context is also possible in a statistical sense \citep{2017MNRAS.465.3622R,2017ApJ...834...69L,2018MNRAS.480.2343C,2019MNRAS.486..331C,2019MNRAS.482.4528E,2019arXiv190902630H} or by including additional semi-analytical modelling \citep{2008MNRAS.387.1131B,2018MNRAS.475.4309P}. 
Recent analytical models favor a "conveyor belt" model \citep{2014prpl.conf..291L} with gas accretion and star formation happening simultaneously \citep{2019arXiv190901565K} which is supported by the analysis of the most massive clusters forming in our simulation \citep{2019ApJ...879L..18L}.  

In this paper we present a dwarf galaxy merger simulation with gas-rich initial conditions, which provide an ideal 
environment for efficient star and cluster formation. Crucially, the simulation has a minimum gas particle mass of $4\, M_{\odot}$ and realizes individual massive stars with their individual tracks and models their radiation and supernova feedback at sub-parsec spatial resolution. This enables us to study the formation and evolution of the stellar cluster population from globular cluster masses down to almost the smallest observed cluster masses of $\sim 200 M_{\odot}$ in a starbursting galactic environment. In an earlier study \citeauthor{2019ApJ...879L..18L} (2019, L19 hereafter)
we studied the properties of massive, globular cluster-like objects, which populate the high-mass end of the cluster mass function. In the present study we concentrate on describing the entire young cluster population formed during the merger and show that our simulated clusters form rapidly on timescales of a few Myrs with properties similar to the observed local cluster population. 

The simulation is a part of the \textsc{griffin} project, which is an acronym for Galaxy Realizations Including Feedback From INdividual massive stars. The aim of this project is to perform galaxy scale simulations of individual galaxies, galaxy mergers, and cosmological zoom simulations at such high resolution and physical fidelity that individual massive stars can be realised and important feedback processes such as supernova explosions \citep{2019arXiv190713153S} can be reliably included to study the formation of a realistic non-equilibrium multi-phase interstellar medium \citep{2017MNRAS.471.2151H}. This level of detail in modern simulations is very important as the environmental density of supernova explosions is controlled by stellar feedback processes and to a large extent by stellar clustering \citep[see][for a detailed discussion of this challenge]{2017ARA&A..55...59N}. A numerical model reliably representing the fundamental mass unit of single massive stars and the fundamental energy injection unit of individual supernova explosions in realistic star cluster populations provides unique insights into the physical mechanisms regulating the multi-phase structure of the galactic ISM as well as the driving of galactic outflows - and therefore galaxy evolution as a whole \citep{2017ARA&A..55...59N}. In this paper we use the high dynamic fidelity of a dwarf merger simulation to study the formation of galactic populations of star clusters across environments changing by many orders of magnitude in density. We show that our model, which has a self-consistently evolving multi-phase interstellar medium, produces CMFs similar to observations over four orders of magnitude in star cluster masses with the most massive systems being realistic proto-globular clusters \citep{2019ApJ...879L..18L}.

This article is organized as follows. The main aspects of the simulation code and initial conditions are described in Section \ref{section:simulations}. A general overview of the simulation, including the star formation history is presented in 
Section \ref{section:general}. This section also introduces our stellar cluster identification procedure and discusses the cluster formation during the simulation. The properties of the young cluster population are analysed in detail in Section \ref{section:star_clusters}, with the results also being compared to the locally observed stellar cluster population. Finally, 
we present our conclusions in Section \ref{section:conclusions}.

\section{Simulations}\label{section:simulations}

\subsection{The simulation code}

The simulations were run with a modified version of the well-tested smoothed particle hydrodynamics (SPH, \citealt{1977AJ.....82.1013L, 1977MNRAS.181..375G}) tree code \gadget\ \citep{2005MNRAS.364.1105S}. Gas dynamics is modelled with the SPH implementation \sphgal, presented in \citet{2014MNRAS.443.1173H, 2016MNRAS.458.3528H, 2017MNRAS.471.2151H}, using the pressure-energy formulation \citep{2010MNRAS.405.1513R, 2013MNRAS.428.2840H, 2013ApJ...768...44S}. The gas properties are smoothed over $100$ neighboring particles  using the Wendland $C^4$ kernel \citep{2012MNRAS.425.1068D}. To stabilize the SPH scheme, we also model artificial viscosity \citep{2010MNRAS.408..669C} with a few important modifications (see \citealt{2014MNRAS.443.1173H}) and artificial conduction of thermal energy \citep{2008JCoPh.22710040P, 2010MNRAS.405.1513R} in converging gas flows. The time stepping is regulated with a limiter which keeps neighboring particles within a time step difference by a factor of four to capture shocks accurately. All technical details are given in \citet{2014MNRAS.443.1173H, 2016MNRAS.458.3528H, 2017MNRAS.471.2151H}. We briefly review the implementations including updates to the star formation model in section \ref{sf}. 

\subsection{Chemistry and cooling}

We track the chemical composition of gas and stars by following the abundance of $12$ elements
(H, He, N, C, O, Si, Mg, Fe, S, Ca, Ne and Zn) and six chemical species (H$_2$, H$^+$, H, CO, C$^+$, O) as well as the free electron density. Dust, which constitutes $0.1\%$ of the gaseous mass at the adopted $\sim 0.1\,Z_\odot$ metallicity, contributes to H$_2$ formation and also the shielding of radiation in the gas clouds. Dust is assumed to be in thermal equilibrium where the dust temperature is calculated from the balance of cooling and heating as described in \citet{2012MNRAS.421..116G}. The dust heating processes included here are stellar emission (the interstellar radiation field), H$_2$ formation on dust, and dust-gas collisions.

The cooling rates for gas are modelled in two temperature regimes. Below $T=3\times10^4$ K we use the chemical network described in detail in \citet{2016MNRAS.458.3528H}, based on \citet{1997ApJ...482..796N}, \citet{ 2007ApJS..169..239G} and \citet{2012MNRAS.421..116G}, to follow low-temperature cooling down to $T=10$ K. High-temperature cooling above $T=3\times10^4\,\rm K$ is modelled using the metallicity-dependent cooling tables from \citet{2009MNRAS.393...99W} assuming an optically thin ISM in an ionizing UV-background from \citet{1996ApJ...461...20H}.

\subsection{Star formation}\label{section:star_formation}
\label{sf}

The onset of gravitational collapse for gas at the local SPH averaged density $\rho_i$ can be approximated with the local Jeans mass, defined as
\begin{equation}
 M_{J,i} = \frac{\pi^{5/2}c_{s,i}^3}{6G^{3/2}\rho_i^{1/2}}
\end{equation}\label{eq:jeans_mass}
where $c_{s,i}$ is the local sound speed and $G$ is the gravitational constant. The gas is allowed to form stars 
stochastically at star formation (SF) efficiency $\epsilon_\mathrm{SF}=2\%$ if the local Jeans mass is less than $8$ SPH kernel masses, which corresponds to $\sim 3200\, M_\odot$. We also require a gas particle to be in a converging flow to form stars. The probability for a gas particle, which meets these criteria, to turn into a stellar particle is then set as $1-\exp(-p)=1-\exp(-\epsilon_\mathrm{SF}\Delta t/ t_{\rm ff})$, where $t_{\rm ff}$ is the local free-fall time $t_{\rm ff}\sim (4\pi G \rho_i)^{-1/2}$ given by the SPH quantities, and $\Delta t$ is the length of the time-step \citep{2017MNRAS.471.2151H}.

In addition to \citet{2017MNRAS.471.2151H}, we enforce star formation if the Jeans mass is resolved by less than 0.5 SPH kernel masses ($\sim 200\, M_\odot$). Any gas particle ending their time step below this threshold is instantaneously turned into a star particle, ignoring the converging flow criterion. The instantaneous star formation limit corresponds to hydrogen number densities of $n_{\rm H}\gtrsim 10^{3.5} \ \rm cm^{-3}$ and temperatures of $T=10 - 100 \ \rm K$. For phase diagrams illustrating the gas properties and the SF thresholds, see Fig. 1 in L19.

\subsection{IMF sampling}

We sample the mass of a star-forming gas particle into an array of stellar masses by drawing randomly from a Kroupa initial mass function (IMF) \citep{2001MNRAS.322..231K}. This enables us to resolve massive stars with individual stellar particles: gas particles which draw a stellar mass greater than the gas mass resolution of $\sim 4\,M_\odot$ represent individual massive stars. Lower mass stars are stored as stellar populations with total mass equal to the mass of the original gas particle. Whenever the total sampled mass exceeds the mass of the original gas particle, the exceeding mass is reduced from the next star forming gas particles in order to conserve mass. Each new stellar particle also inherits the chemical composition of their progenitor gas particle.

\subsection{The Interstellar radiation field}\label{sec:ISRF}

We model the spatially and temporally evolving interstellar radiation field (ISRF) emanating from stellar particles which include sampled stars younger than their lifetimes. The mass-dependent stellar lifetimes are calculated from \citet{2013A&A...558A.103G} assuming a metallicity of $Z=0.002\approx 0.1\, Z_\odot$. The stellar particles emit a far-ultraviolet (FUV) radiation field out to a radius of $50$ pc for which we assume an optically thin medium, appropriate for our low-metallicity system. The stellar FUV spectrum is integrated from the \basel\ library \citep{1997A&AS..125..229L, 1998A&AS..130...65L, 2002A&A...381..524W} in the range $6$--$13.6$ eV, and summed up for all (young enough) stars in each stellar particle.  We account for dust extinction and shielding by the chemical species using the \treecol\ \citep{2012MNRAS.420..745C} algorithm along 12 line-of-sight columns obtained using \healpix\ \citep{2011ascl.soft07018G}.

We also propagate the hydrogen-ionizing radiation from massive stars assuming a balance between the production rate of ionizing photons and the recombination rate in the ISM surrounding the stars. The photoionization implementation produces successfully analytical solutions to D-type expansion fronts \citep{1978ppim.book.....S, 2015MNRAS.453.1324B} and it can cope with overlapping HII regions around young massive stars. The impact of the ionizing radiation on the ISM is important for the accurate modelling of the following supernova events, as the massive stars heat up their surroundings before they explode in a lower-density medium, as opposed to the stellar birth cloud \citep{2015MNRAS.449.1057G}. 

\begin{figure}
\includegraphics[width=\columnwidth]{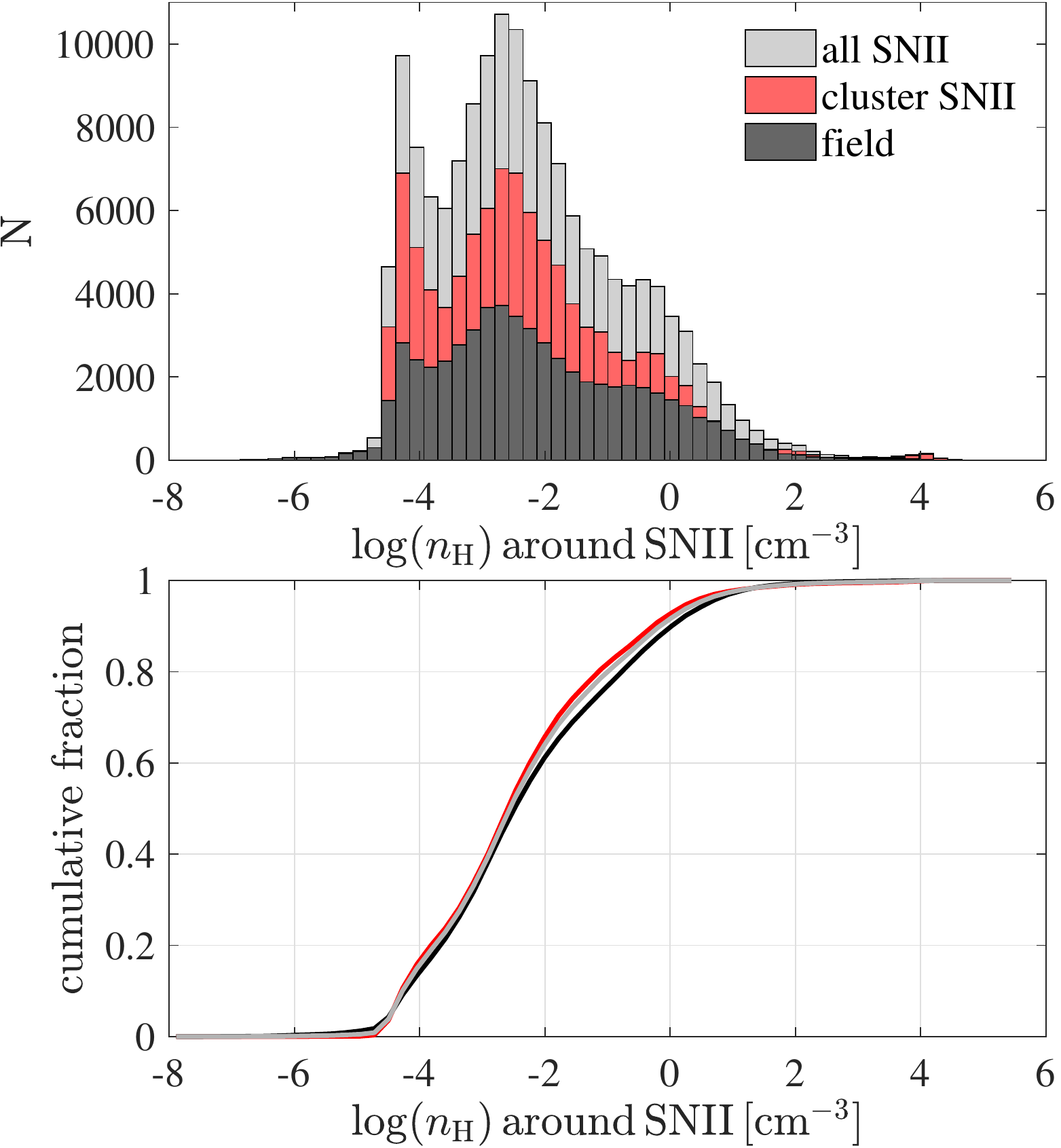}
\caption{Distribution of all environmental densities of SNII during the simulation up to $100$ Myr after the starburst. The SNII events have been separated according to whether the SN remnant is a star cluster member. About $97 \%$ of the SNII explode at densities $\lesssim 10 \ \rm cm^{-3}$ and can be considered well resolved  at this resolution. Less than $1 \%$ of the SN explode at densities of  $\gtrsim 100 \ \rm cm^{-3}$ and have marginally or un-resolved hot phase generation. Still, the radial momentum injection is typically captured \citep{2016MNRAS.458.3528H,2019arXiv190713153S}. \label{fig:SN_environment}}
\end{figure}

\subsection{Stellar feedback}

Each individual star with a mass greater than $8\,M_\odot$ will explode as a type II supernova (SNII) at the end of its stellar lifetime \citep{2013A&A...558A.103G}.
At our mass resolution we resolve the expansion of the supernova remnants self-consistently with stochastic energy injection (see Appendix B in \citealt{2016MNRAS.458.3528H}), where each SNII releases $10^{51}$ erg as thermal energy into 100 nearest gas particles weighted by a cubic spline kernel. In addition, the SNII events release metals according to the metallicity-dependent ejecta release rates obtained from \citet{2004ApJ...608..405C} at an ejection velocity of $3000$ km$/$s. Stars also release AGB (asymptotic giant branch) winds at a more gradual rate and lower velocity of $25$ km$/$s, with metal yields obtained from \citet{2010MNRAS.403.1413K}.

Because of stellar motion out of the dense birth places, stellar clustering, and photoionization \citep{2017MNRAS.471.2151H}, the vast majority of the SNII events go off in a considerably less dense gaseous environment than e.g. the densities at which the star clusters form \cite[see e.g.][]{2016MNRAS.456.3432G,2017MNRAS.466.3293P, 2017MNRAS.471.2151H, 2017ARA&A..55...59N}. In Fig. \ref{fig:SN_environment} we show a histogram of the ISM densities around all SNII events that occur until 100 Myr past the starburst. In the top panel of Fig. \ref{fig:SN_environment} we separate the supernova remnants which become part of star clusters (see Section \ref{section:cluster_identification}) and the field remnants, and show the cumulative distribution in the bottom panel.  Over $97\%$ of the SNII  explode at ISM number densities less than $n_\mathrm{H}=10\, \mathrm{cm}^{-3}$ and can be considered well resolved \citep{2016MNRAS.458.3528H}. Only $1\%$ of all the SNII explode at a density above $n_\mathrm{H}=100\, \mathrm{cm}^{-3}$, out of which more than $70\%$ occur in clusters. The peak of the density distribution is at $\sim 10^{-2}\, \rm cm^{-3}$ similar to other high resolution simulations with structured multi-phase ISM properties \citep{2016MNRAS.456.3432G, 2017MNRAS.466.3293P, 2017MNRAS.471.2151H}.  

The kernel weighting of the feedback ejecta favors the nearest particles, and at times single gas particles may receive up to a few times their original mass worth of stellar material when for example a nearby massive star explodes as a type II supernova. Whenever a gas particle exceeds a mass limit of twice the original gas mass resolution (here $4\,M_\odot$), its mass is split into two new particles. The ability of the code to resolve supernova blast waves in different environments has been tested in detail in \citet{2017MNRAS.471.2151H} and \citet{2019arXiv190713153S}. 
\begin{figure*}
\includegraphics[width=0.95\textwidth]{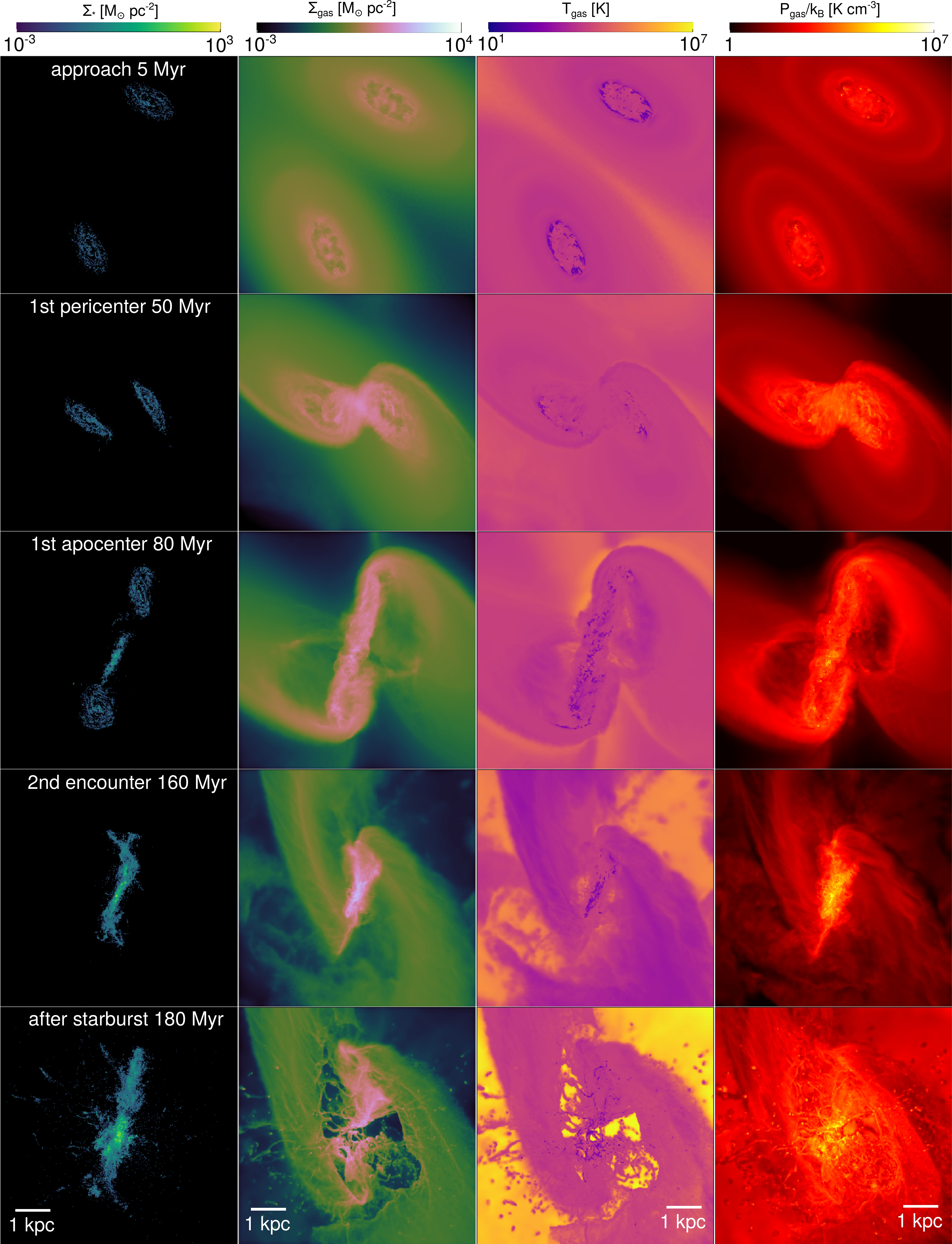}
\caption{An overview of the merger. The rows from top to bottom depict a projection in the $x$--$y$ plane at the approach, the first pericentric passage, the first apocenter, the onset of the starburst, and the interaction slightly after the starburst. The columns from left to right show the surface density of stars formed during the simulation, the surface density of gas, the density weighted gas temperature, and the density weighted thermal gas pressure. The panels span a $(7\,\rm{kpc})^3$ box. \label{fig:overview}}
\end{figure*}

\subsection{Initial conditions}

The initial conditions are based on the dwarf galaxy model used in the simulation studies performed
in \citet{2016MNRAS.458.3528H, 2017MNRAS.471.2151H}. 
The two identical dwarf galaxies with virial masses of $M_\mathrm{vir}=2\times10^{10}\,M_\odot$ and virial radii of $r_\mathrm{vir}=44$ kpc are both composed of a dark matter halo and a gas-rich disk with a rotationally supported stellar population. The dark matter halo is represented with
a Hernquist density profile \citep{1990ApJ...356..359H} with an NFW-equivalent concentration parameter of $c=10$
\citep{2005MNRAS.361..776S} and spin parameter of $\lambda=0.03$ \citep{2001ApJ...555..240B}. 

The exponential stellar disk with a mass of $M_*=2\times10^7\,M_\odot$ is set up with scale radius of $r_{\rm *}=0.73$ kpc and scale height of $h_*=0.35$ kpc. The gaseous disk has initially a mass of $M_{\rm gas}=4\times10^7\,M_\odot$ and scale radius twice that of the stellar disk, $r_{\rm gas}=1.46$ kpc, and the vertical structure is calculated assuming hydrostatic equilibrium. The gaseous metallicity is initially set to $Z_{\rm gas}=0.002\approx0.1\,Z_\odot$, which is a typical value for low-mass dwarf galaxies. 

Each dwarf galaxy consists of $4$ million dark matter particles, $10$ million gas particles and $5$ million stellar particles,
defining a particle mass resolution of $m_{\rm DM}=7\times 10^3\,M_\odot$ and $m_{\rm bar}=4\,M_\odot$ for the dark matter and baryonic components, respectively. The gravitational softening lengths are set to $\epsilon_{\rm DM}=62$ pc for dark matter and $\epsilon_{\rm bar}=0.1$ pc for baryonic particles. The two dwarf galaxies are set on parabolic orbits with a pericentric distance of $d_{\rm peri}=1.46$ kpc and an initial separation of $d_{\rm init}=5$ kpc, leading to a fairly rapid coalescence time for the gas disks. The inclination and argument of pericenter are chosen off-plane as $\{i_{1}, i_{2}\}=\{60^{\circ}, 60^{\circ}\}$ and $\{\omega_{1}, \omega_{2}\}=\{30^{\circ}, 60^{\circ}\}$. We chose the same initial orientations as for the Antennae-like simulations studied in \citet{2018MNRAS.475.3934L}.

As presented in L19, we have run the simulation for up to 100 Myr after the merger, for a total of $\sim 300\,\rm Myr$. Here we concentrate on the first 200 Myr of the simulation time, up to some $40$ Myr past the starburst.
To assess the effects specific to the merging dwarf galaxies, we also run one dwarf galaxy in isolation for the same total simulation duration. More detailed discussion on the properties of the isolated dwarf can be found in \citet{2016MNRAS.458.3528H, 2017MNRAS.471.2151H}.

\section{Star formation in a gas-rich dwarf galaxy merger}\label{section:general}

\subsection{Simulation overview}

A general overview of the evolution of the stellar and gaseous components is shown in Fig. \ref{fig:overview} for five epochs during the simulation: the approach at $5$ Myr, the first pericentric passage at $50$ Myr, the first apocenter at $80$ Myr, the second encounter and the onset of the starburst at $160$ Myr, and at $180$ Myr, slightly past the most intense starburst.

The dwarfs begin to show signs of the interaction first in the outskirts of their extended gas disks. The interaction-induced gas compression is already strong further away from the galactic centers, thus the distribution of star formation is relatively more extended compared to the interactions between more massive late-type galaxies that show more centrally concentrated star formation (e.g. \citealt{1996ARA&A..34..749S,2007AJ....133..791S,2013MNRAS.435.3627E}). During the first encounter and up to a simulation time of $100$ Myr, the tidal bridge dominates the star formation over the combined dwarf disks (see the third row in Fig. \ref{fig:overview}). During the second passage and the starburst phase there are multiple clumpy star formation regions which are distributed mostly in a region $2$ kpc across, comparable to the scale radii of the initial disks. Most of the star formation is also located half a kiloparsec off-center from the dark matter distribution. When the SFR peaks, the gas disks are settling into the central region, where the most massive stellar clusters are forming. After the starburst, the supernova feedback from the stellar clusters unbinds the gas even in the densest star-forming regions, dispersing the gas and halting the most intense starburst.  


\begin{figure*}
\includegraphics[width=\textwidth]{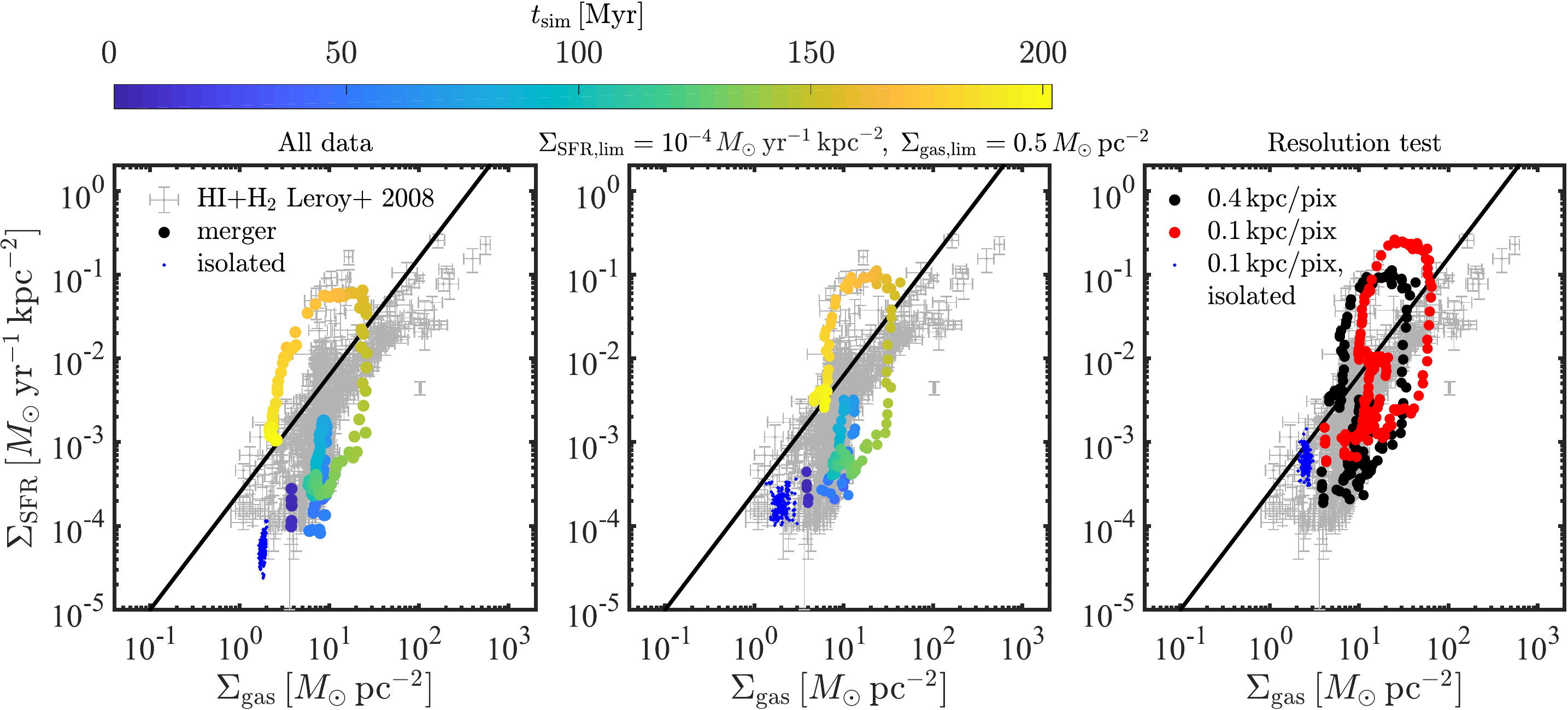}
\caption{The evolution of the projected SF properties in the merger and the isolated galaxy. The leftmost panel shows the full dataset without any observational restrictions averaged over pixels with SF, at a resolution of $400$ pc per pixel. In the middle panel the average over gas surface density and SFR surface density have been limited to pixels exceeding the observational sensitivity limits of $\Sigma_\mathrm{gas,lim}=0.5\, M_\odot\, \mathrm{pc}^{-2}$ and $\Sigma_\mathrm{SFR,lim}=10^{-4}\, M_\odot\,\mathrm{yr}^{-1}\, \mathrm{kpc}^{-2}$, analogous to the sensitivity in the THINGS survey \citep{2008AJ....136.2782L}. The middle and right panels additionally only consider pixels including IMF sampled stars above $8\,M_\odot$, to mimic observing UV and H$_\alpha$ emission. The rightmost panel shows a resolution study for the data in the middle panel. Observed galaxies from \citet{2008AJ....136.2782L} are shown on the background and the diagonal line shows the traditional Kennicutt-Schmidt \citep{1998ApJ...498..541K} relation of $\bold{ \Sigma_\mathrm{SFR}=2.5\times 10^{-4} \Sigma_\mathrm{gas}^{1.4}}$. \label{fig:ks_gas}}
\end{figure*}

\begin{figure*}
\includegraphics[width=\textwidth]{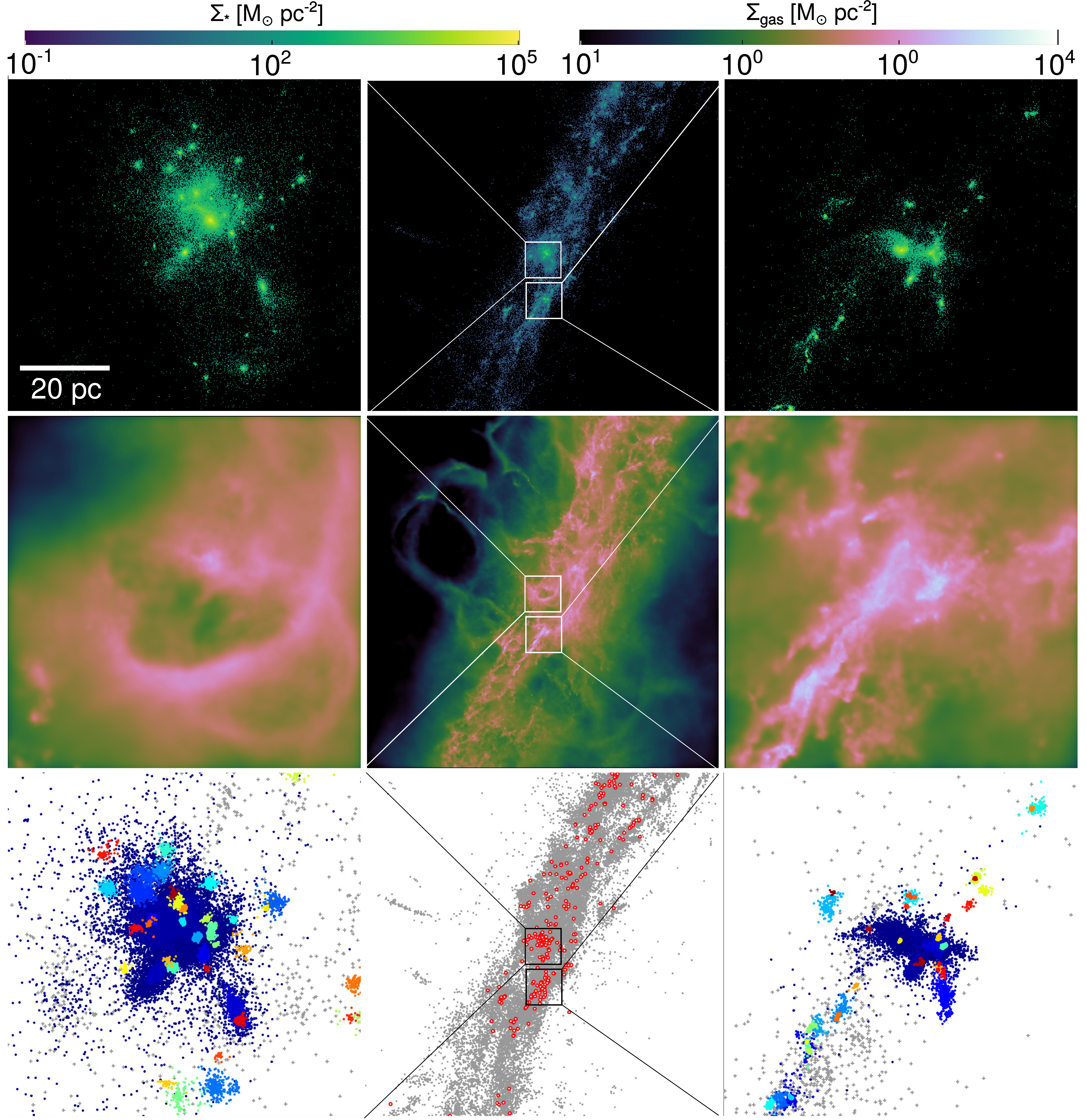}
\caption{An $x$--$y$ projection of the stellar and gaseous structure around the central cluster formation region at a simulation time of $t \sim160$ Myr, before the first massive cluster blows out the super-bubble. The top and middle rows show the stellar surface density and gas surface density around the two most massive star formation region at that time. The left and right bottom panels show the bound cluster particles (colored points) and unbound stellar particles (gray) within the respective regions. All cluster positions are shown with red points in the middle panel of the bottom row. The left and right columns show $80$ pc by $80$ pc by $80$ pc zoomed slices of the regions shown in the middle column, which spans a $800$ pc by $800$ pc by $800$ pc slice centered at the first forming massive cluster.}
\label{fig:cluster_overview}
\end{figure*}

\subsection{The global star formation rate}

The total SFR increases to a few times $10^{-3}\, M_\odot\,\rm{yr}^{-1}$ during the first pericentric passage. During the starburst, the SFR peaks at $\sim 0.2$--$0.3\,M_\odot\, \mathrm{yr}^{-1}$ at a simulation time of $t=160-170$ Myr, which corresponds to a specific SFR of $\rm sSFR=4.6$--$7\times 10^{-9}\, \mathrm{yr}^{-1}$ at the total stellar mass of $\sim4.3\times 10^7\,M_\odot$. At its highest, the specific SFR is therefore up to an order of magnitude above the general star-forming main sequence \citep{2007ApJS..173..267S, 2007ApJS..173..315S,2012AJ....143..133H}, although the observed scatter in the relation is quite large for low mass galaxies \citep{2009MNRAS.400..154B, 2016MNRAS.459.1827P}. The SFR in our merger is similar to interacting dwarf galaxies in general \citep{2016MNRAS.459.1827P, 2018ApJS..237...36P}. In the context of other numerical studies, for example the dwarf merger system presented in \citet{2017ApJ...846...74P} has a stellar mass of $\sim 9\times 10^7\,M_\odot$ and a total SFR of a few $0.1\,M_\odot\, \mathrm{yr}^{-1}$.

The total star formation efficiency, calculated as all stellar mass formed during the simulation divided by the original gas mass, reaches $6.3\%$ by the end of the starburst. With respect to the pre-existing stellar disk, the newly formed stars account for $11\%$ of all stellar matter in the system.

\subsection{Resolved $\Sigma_\mathrm{SFR}$--$\Sigma_\mathrm{gas}$-relation}\label{section:KS_relation}

The evolution of the star formation rate surface density $\Sigma_\mathrm{SFR}$ with respect to gas surface density $\Sigma_\mathrm{gas}$ is shown in Fig. \ref{fig:ks_gas} in $1$ Myr steps for both the merger and the isolated dwarf galaxy. The evolution along the merger is color coded by time. We calculate the resolved values by default in $400$ pc by $400$ pc pixels, and only include in our calculation pixels with ongoing star formation, akin to the SFR weighted $\Sigma_\mathrm{SFR}$ in e.g. \citet{2017ApJ...839...78J}. 

We also model the effects of observational restrictions by calculating the $\Sigma_\mathrm{SFR}$ and $\Sigma_\mathrm{gas}$ values from both the full dataset and by applying the observational sensitivity limits of the THINGS survey \citep{2008AJ....136.2782L}, which provides spatially resolved measurements at sub-kpc resolution. Firstly, we only include pixels with stellar particles where the IMF sampling has produced a star more massive than $8\,M_\odot$ when we calculate the SFR surface density. This mimics observing the far-UV and re-emitted $24\,\mu\mathrm{m}$ dust emission originating from O/B stars. Secondly, we apply sensitivity limits to both the SFR surface density, $\Sigma_\mathrm{SFR,lim}=10^{-4}\, M_\odot\, \mathrm{kpc}^{-2}$, and the gas surface density, $\Sigma_\mathrm{gas,lim}=0.5\, M_\odot\, \mathrm{pc}^{-2}$, which ignore the regions with too weak signal to be detected. The observational limits reduce the number of detected pixels from the typical values of $50$--$200$ by $30$--$60\%$ during the first and second passage and the starburst phase, and by up to $\sim70\%$ thereafter.

The leftmost panel of Fig. \ref{fig:ks_gas} shows the relationship between the gas density and star formation rate surface density without the application of any observational limitations, whereas the middle and right side panels include the observational restrictions detailed above. The final result for each snapshot is an average over all the pixels with a detectable signal. We also test the effect of better image resolution in the rightmost panel of Fig. \ref{fig:ks_gas}. The obtained values are compared to the global Kennicutt-Schmidt star formation relation of the form $\Sigma_\mathrm{SFR}\propto \Sigma_\mathrm{gas}^{1.4}$ \citep{1998ApJ...498..541K} and the observed $\Sigma_\mathrm{SFR}$--$\Sigma_\mathrm{gas}$ values from the THINGS survey \citep{2008AJ....136.2782L} which has a pixel resolution of $400$ pc$/$pix in dwarfs and $800$ pc$/$pix in other galaxies.

The isolated dwarf is located in the low $\Sigma_\mathrm{SFR}$ -- $\Sigma_\mathrm{gas}$ region of the observed dataset, as could be expected for a quiescent dwarf galaxy. In the middle panel of Fig. \ref{fig:ks_gas} the mean $\Sigma_\mathrm{SFR}$ increases to the imposed $\Sigma_\mathrm{SFR}$ limit when pixels with only a few new stars and therefore not enough signal, are ignored. The gas distribution is fairly uniform at 400 pc per pixel scale, therefore the limit on $\Sigma_\mathrm{gas}$ does not affect the mean results for $\Sigma_\mathrm{gas}$. For a resolved, more detailed study of the star formation properties of the isolated dwarf, see \citet{2016MNRAS.458.3528H}. 

The merger starts from the same region as the isolated dwarf, and evolves through most of the entire range of observed data during the simulation. The observational restrictions in the middle panel shift the simulated data slightly up along the traditional KS-relation, as for example blown out gas and small isolated star formation regions are excluded from the global mean. The post-starburst data points of the merger shift to higher gas surface densities because the clumpy star formation produces high-mass stars more frequently in high-density regions, which are less severely affected by the observational restrictions imposed in the middle panel of Fig. \ref{fig:ks_gas}.

The rightmost panel of Fig. \ref{fig:ks_gas} shows for comparison the data from the middle panel calculated at a resolution of $100$ pc per pixel. Increasing the spatial resolution shifts the measured data further up along the KS-relation as long the resolution is coarser than the scale of single molecular clouds, where observations report a steepening of the star formation relation \citep{2011ApJ...739...84G, 2015ApJ...809...87W}. In the isolated galaxy and the early merger phases, the uniform gas distribution results in unaffected values for $\Sigma_\mathrm{gas}$ while increasing resolution helps in separating regions with and without star formation.

\subsection{Identification of stellar clusters}\label{section:cluster_identification}

We use the built-in friends-of-friends (FoF) and \subfind\ \citep{2001MNRAS.328..726S, 2009MNRAS.399..497D} algorithms of \gadget\ to identify bound stellar clusters during the simulations. The FoF group finding and the \subfind-unbinding are performed only for new stellar particles formed during the simulations. In detail, \subfind\ differentiates 
all bound clusters in each snapshot by methodically examining all density peaks even if they are embedded in a more extended object.
Our method differs from observational techniques in that we use the full 3D data available from the simulation, while observational data reduction often suffers from the loss of information due to projection effects. We however chose to employ \subfind\ for our cluster identification, as one of our goals is to study the build-up and merging of the young clusters, which might at different stages of their evolution be even temporarily embedded within each other, especially when seen in projection.

In defining a stellar cluster, we use a minimum particle number of $50$, which translates into a minimum cluster mass of $M_{\mathrm{cl,min}}\sim 200\,M_\odot$ at our resolution. Our choice is somewhat more conservative compared to e.g. the definition of a minimum of $35$ stars for a star cluster in \citet{2003ARA&A..41...57L}. On the other hand, observational samples of stellar clusters outside of the Milky Way are typically only complete above a few hundred $M_\odot$ in the  M31 \citep{2015ApJ...802..127J} and the Magellanic clouds \citep{2003AJ....126.1836H} and a few $10^3\,M_\odot$ outside of the Local Group \citep{2012ApJ...751..100C, 2018MNRAS.473..996M}. When studying the individual clusters, we will consider the entire simulated sample, whereas when fitting the cluster mass functions we will restrict the data in correspondence to the observed completeness limits.

An overview of the central cluster formation environment is shown in Fig. \ref{fig:cluster_overview}, in a $800$ pc box centered on the most massive cluster formation region. The snapshot shown here is from the onset of the starburst at a simulation time of $t=160$ Myr, when the destructive feedback from the young stellar population has not yet destroyed the gaseous structures in the star-forming central region (see Fig. \ref{fig:overview}). The left and rightmost panels in the top and middle row of Fig. \ref{fig:cluster_overview} show the distribution of stars and gas in two $80$ pc zooms around the most massive cluster formation regions, which will give birth to the GC-like massive clusters. Stellar feedback in the leftmost panel of Fig. \ref{fig:cluster_overview} has already started to form a cavity around the young stellar clusters, whereas the rightmost panel shows smaller mass clusters still embedded in the gaseous filament in which they form.

The left and rightmost panels in the bottom row of Fig. \ref{fig:cluster_overview} show with separate colors the stellar particles bound to each individual cluster identified by \subfind\ from the  $80$ pc slices in the top row. Most of the stellar particles in these two star-forming regions not bound to smaller objects are considered bound to the most massive cluster in each region, which is also why the masses and formation rates reported here are upper limits. Many of the smaller mass proto-clusters, each with their own bound population of stars, seem embedded in the massive clusters at least seen in projection. Some of these low mass clusters will coalesce with the more massive clusters, some will get disrupted, and the rest will remain as young individual star clusters. Plenty of clearly separate smaller mass clusters are already present, especially in the beads-on-a-string structures such as the bottom-left region of the rightmost column of Fig. \ref{fig:cluster_overview}. 

The middle panel of the bottom row of Fig. \ref{fig:cluster_overview} shows the center of mass of each cluster in the central region (top middle panel). The spatial distribution of clusters is mostly filamentary both on large and small scales, where the star and cluster formation follows the structure of the gas distribution.

\begin{figure}
\includegraphics[width=\columnwidth]{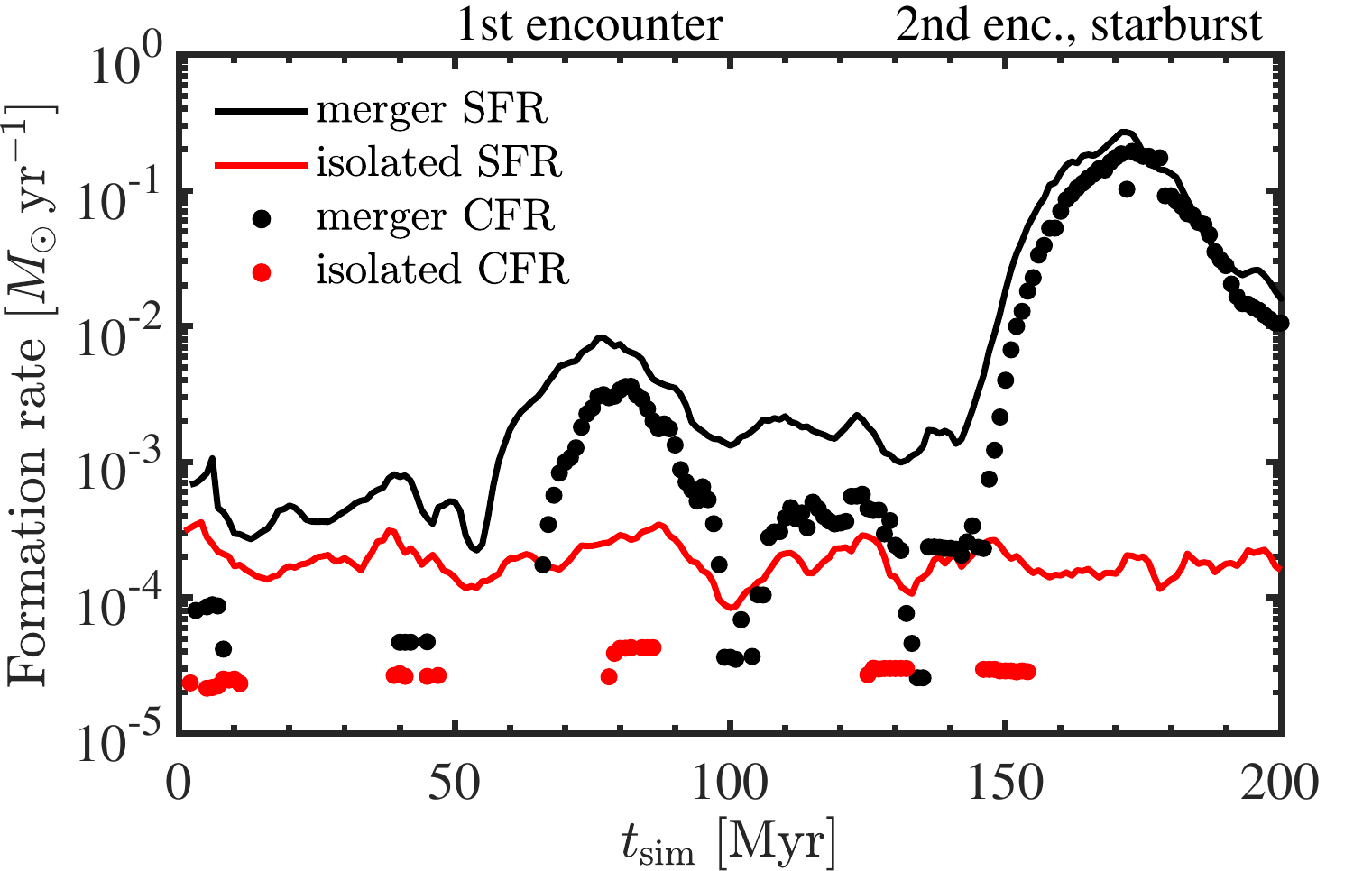}
\caption{The total star formation rates of the merger and the isolated dwarf galaxy calculated for each snapshot from stellar particles with ages less than $5$ Myr. The SFR is compared to the upper limit of the cluster formation rates, computed from clusters younger than $10$ Myr. The snapshots with no young clusters have been omitted, which results in gaps in the data, especially for the isolated dwarf. \label{fig:cfr}}
\end{figure}

\subsection{Cluster formation rate}

The evolution of the merger can be divided into three phases where the increase in star formation and the mass in young stellar clusters trace each other. In Fig. \ref{fig:cfr} we show the SFRs of the merger and the isolated dwarf compared to the cluster formation rates (CFRs) in both simulations defined as
\begin{equation}\label{eq:cfr}
 \mathrm{CFR} = \frac{M_\mathrm{*,\, cl}(\tau <10\, \mathrm{Myr})}{10\, \mathrm{Myr}}
\end{equation}
where $\tau$ is the mean stellar age and $M_\mathrm{*,\, cl}(\tau <10 \mathrm{Myr})$ is the stellar mass in clusters with $\tau<10$ Myr. We chose quite a short timescale of $10$ Myr for our analysis, compared to e.g. studies of quiescent disk galaxies such as in \citet{2016ApJ...827...33J}, as the changes in the SFR and the subsequent CFR during the merger occur quite rapidly. In contrast to observational CFR results where the rates are lower limits, the CFR presented in Fig. \ref{fig:cfr} is an upper limit since we include here all the cluster data obtained from the simulation without corrections for observational bias. 

As defined in Eq. \ref{eq:cfr}, each cluster appears in the CFR data while it has a mean age less than the threshold age of $10$ Myr and a mass above the lower mass limit of $\sim 200\, M_\odot$.
The quiescent phase of star formation during the approach phase and in the isolated dwarf is accompanied by an episodic CFR where new low mass clusters form in a $40$--$50$ Myr cycle.

In the merger simulation the cluster formation is more continuous. The CFR along the merger increases after the first pericenteric passage, but only up to half of the respective SFR. Most of the stars form outside of the few tens of bound clusters before the second passage at $160$ Myr. After the second passage the mode of star formation abruptly changes, as now unlike in the quiescent phase and during the first passage, most of the star formation takes place in the cluster forming regions of the central high pressure environment (see Fig. \ref{fig:overview}). By simulation time $200\,\rm Myr$, $70\%$ of all the stars formed have ended up in bound clusters, which corresponds to $7.8\%$ of all the stellar mass in the system. 

\begin{figure}
\includegraphics[width=\columnwidth]{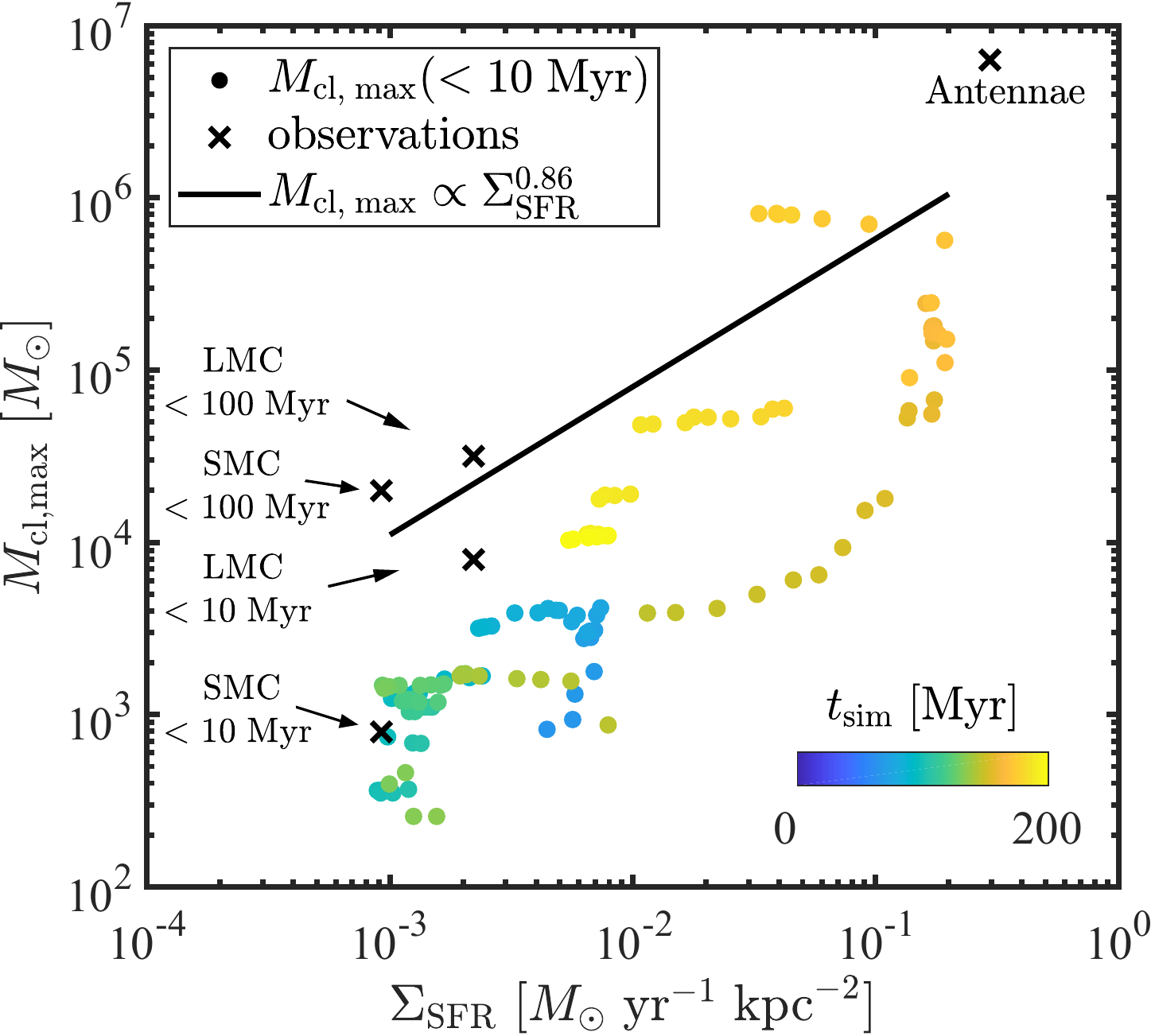}
\caption{The mass of the most massive young cluster ($<10\, \mathrm{Myr}$) as a function of the SFR surface density, calculated in pixels with SF at resolution of $100$ pc per pixel. The black crosses show observed values for the most massive young clusters in the LMC and SMC (two age bins labeled on the left, \citealt{2003AJ....126.1836H}) and in the Antennae merger (cluster age of the order of $1$ Myr, \citealt{2010AJ....140...75W}) where the $\Sigma_\mathrm{SFR}$ values are from \citet{2017ApJ...849..128C} and \citet{2017ApJ...839...78J} respectively. The black diagonal line shows a model relation from \citet{2018ApJ...869..119E}. \label{fig:SFR_Mcl}}
\end{figure}

\begin{figure*}
\includegraphics[width=\textwidth]{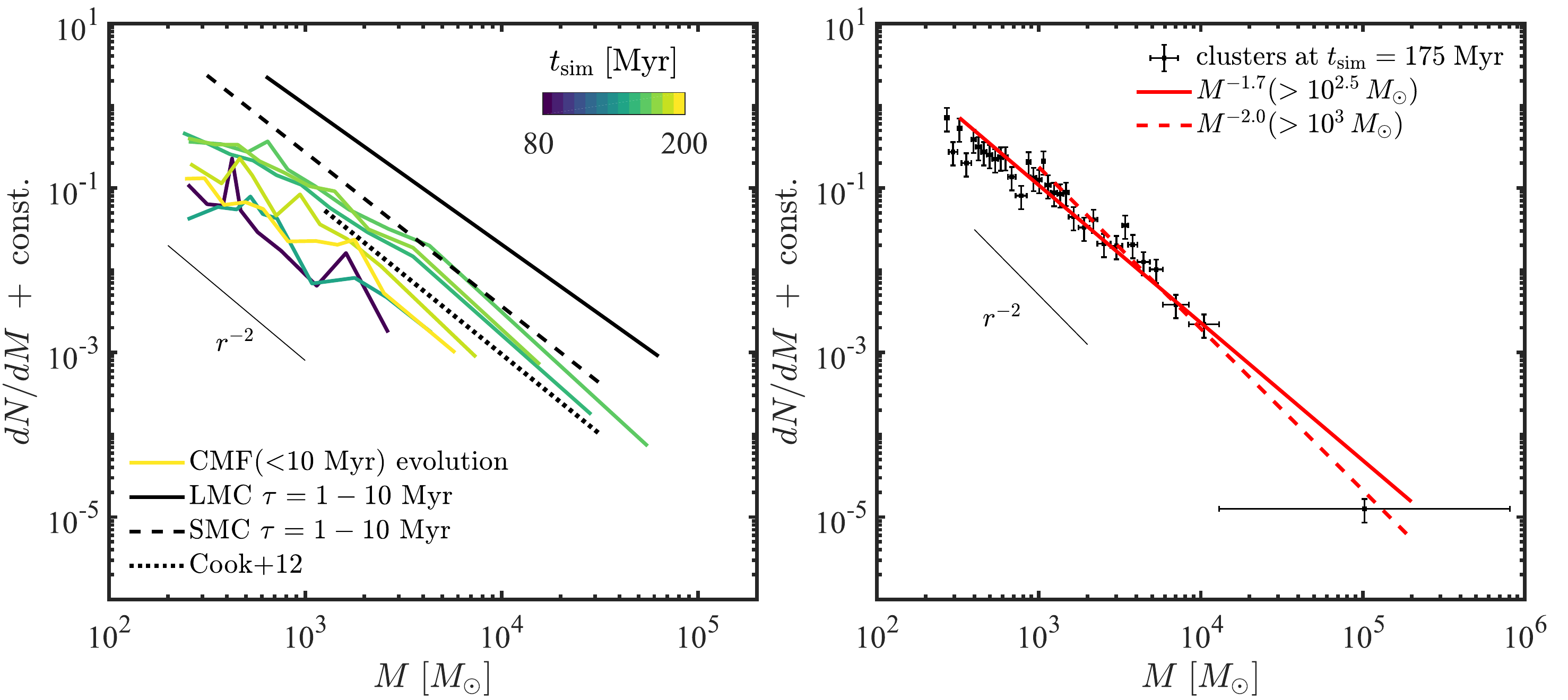}
\caption{Left panel: the CMF of young ($<10$ Myr) clusters after first passage at $80$ Myr and from the second passage until the merger ($150$--$200$ Myr) in $10$ Myr steps. The clusters in each snapshot are binned into $10$ bins with roughly equal number of clusters in each bin, omitting snapshots with less than 2 clusters per bin.
The CMFs in the LMC (solid line) and SMC (dashed line) for clusters with ages between $1$--$10$ Myr are from \citet{2012ApJ...752...96F} and the ANGST survey CMF (dotted line) including clusters with ages up to $63$ Myr is from \citet{2012ApJ...751..100C}. Right panel: same as on the left, immediately after the merger at time $175$ Myr for maximum population size, now binned with 10 clusters per bin and showing the Poisson errors. The best fit CMF to clusters above $10^{2.5}\, M_\odot$ and $10^{3}\, M_\odot$ correspond to power-laws with indices $\alpha=-1.67\pm 0.15$ and $\alpha=-1.96\pm 0.18$, respectively. \label{fig:mass_function}}
\end{figure*}

\subsection{Maximum cluster mass}

The maximum mass of the most massive stellar cluster is thought to be set by the combined disruption from shear forces and stellar feedback, with the final mass depending on the formation environment \citep{2017MNRAS.469.1282R}. Dwarf galaxies and interacting galaxies, with weak rotational shear, are found to harbor the most massive young stellar clusters \citep{2010ApJ...724.1503W}.

As the first step in connecting the cluster formation to the SF we show in Fig. \ref{fig:SFR_Mcl} the most massive young ($<10$ Myr) cluster as a function of the mean star formation rate surface density $(\Sigma_\mathrm{SFR})$ in $100$ pc pixels with ongoing star formation, in each snapshot along the entire merger. The maximum cluster mass follows the evolution of the SFR surface density. We compare our results to two age groups in the LMC and SMC \citep{2003AJ....126.1836H, 2017ApJ...849..128C}, and to the observed most massive super star cluster in the Antennae merger \citep{2010AJ....140...75W, 2017ApJ...839...78J}. The age group of massive clusters younger than $10$ Myr in the LMC and SMC agree well with our clusters formed during the first and second passages. The older age groups in the Magellanic clouds may on the other hand represent a different SF environment present at the time of their formation, uncorrelated with the present-day $\Sigma_\mathrm{SFR}$, thus showing unsurprisingly worse agreement with our results. The most massive clusters forming during the starburst fall along a trend between the young clusters in the Magellanic clouds and the upper extreme set by the Antennae.  We also compare to a model relation for self-gravitating clouds from \citet{2018ApJ...869..119E}, which shows a slope similar ($\beta=0.86$) to our trend.

\section{The population of star clusters}\label{section:star_clusters}

\begin{figure*}
\includegraphics[width=\textwidth]{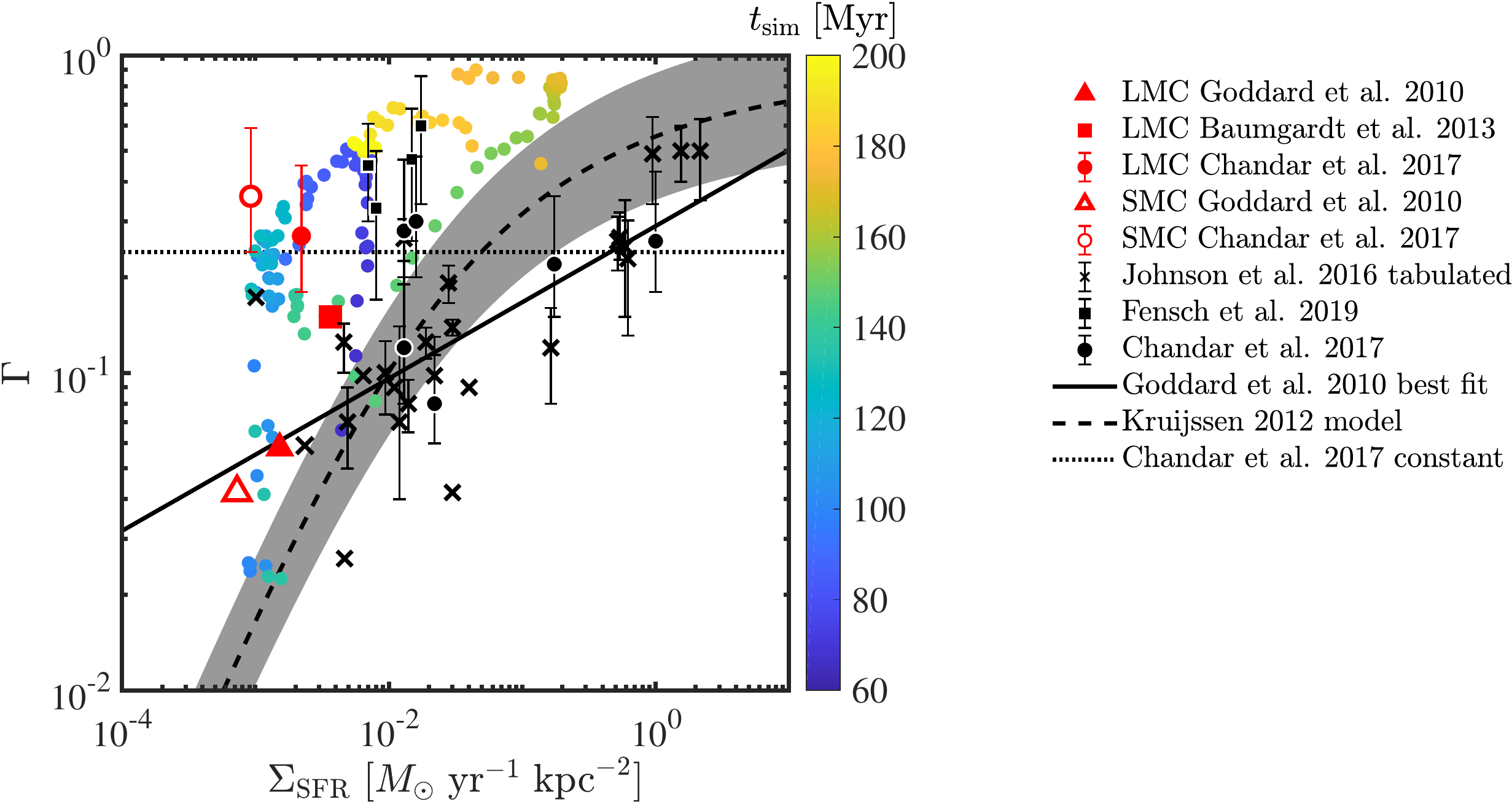}
\caption{Cluster formation efficiency (Eq. \ref{eq:gamma_sfr}) as a function of $\Sigma_\mathrm{SFR}$ from the first pericentric passage onward. The $\Sigma_\mathrm{SFR}$ has been calculated as the mean in pixels with SF, at pixel size of $100$ pc.
The solid line shows the observed relation from \citet{2010MNRAS.405..857G}, the dashed line shows a model relation presented in \citet{2012MNRAS.426.3008K}, with a $\pm 0.2$ dex region indicated in gray, and the dotted horizontal line is the constant $24\%$ CFE found by \citet{2017ApJ...849..128C}. The red markers show observed values for the LMC \citep{2010MNRAS.405..857G, 2013MNRAS.430..676B, 2017ApJ...849..128C} and the SMC \citep{2010MNRAS.405..857G, 2017ApJ...849..128C}. The black squares with errorbars are from \citet{2019A&A...628A..60F}, the black circles from \citet{2017ApJ...849..128C} and the crosses collected from \citet{2010MNRAS.405..857G}, \citet{2011MNRAS.417.1904A}, \citet{2011A&A...529A..25S}, \citet{2011AJ....142..129A}, \citet{2012ApJ...751..100C}, \citet{2014AJ....148...33R}, \citet{2015ApJ...804..123L}, \citet{2015MNRAS.452..246A} and \citet{2016ApJ...827...33J}. Errors for the observed points without errorbars are either unavailable or smaller than the marker size. \label{fig:cfe}}
\end{figure*}

\subsection{Cluster mass function}

The mass function (MF) in terms of the number of objects $N$ per mass bin $dM$ for both interstellar molecular clouds and stellar clusters can be described by a power-law $dN/dM \propto M^{\alpha}$
with a power-law index of $\alpha$. Observationally, molecular clouds have been shown to agree with a power-law index of $\alpha\sim -1.7$ -- $ -1.8$ \citep{1991ApJ...368..432L, 1998A&A...331L..65H, 1998A&A...329..249K}. Similarly, the cluster mass function (CMF) has been observed with a slightly steeper power-law index, closer to $\alpha\sim -2$ (see e.g. \citealt{2003ARA&A..41...57L} and \citealt{2010ARA&A..48..431P} and references therein). 

One of the strong points of our simulation comes from the combination of detailed subresolution models with high spatial and particle mass resolution. As a result we are able to follow the cloud collapse and the subsequent formation of stellar clusters in a wide mass range from a few $100\,M_\odot$ to $10^6\,M_\odot$ during the entire interaction sequence. The evolution of the binned CMF calculated from young clusters ($<10$ Myr) is given in the left panel of Fig. \ref{fig:mass_function} in the dwarf galaxy merger after the first pericenter at $80$ Myr and in $10$ Myr intervals from $150$ Myr onward. The cluster masses have been binned with equal number of clusters per bin, adaptively for each snapshot so that there are always ten bins per snapshot. The overall number of clusters in all of the mass bins increases while the slope of the CMF is established already after the first passage and remains quite constant as the merger proceeds. As shown in Fig. \ref{fig:SFR_Mcl}, the mass of the most massive young cluster increases with increasing $\Sigma_\mathrm{SFR}$, as the CMF shifts upward. The upper limit of the CMF and the hierarchical mass distribution towards lower masses is therefore simply set by the star formation environment, as discussed e.g. in \citet{2015MNRAS.454.1658K} and more recently in \citet{2019ApJ...872...93M} and \citet{2019MNRAS.490.1714P}. 

The CMFs of the young clusters are compared in Fig. \ref{fig:mass_function} against the power-law fits with slopes of $-1.7\pm 0.02$ and $-1.87\pm 0.25$ to young stellar clusters in the Large (LMC) and the Small Magellanic Clouds (SMC) respectively \citep{2012ApJ...752...96F}. Fig. \ref{fig:mass_function} includes also the CMF with a slope of $-1.94\pm 0.26$ fit to clusters in dwarf galaxies from the ANGST project \citep{2012ApJ...751..100C}, fit above their completeness limit of $\sim 10^{3.5}\,M_\odot$ for clusters with ages up to $63$ Myr. Our results for the simulated CMF shapes show good agreement with the observed CMFs. Note that the normalization of the CMFs depends on the total number of clusters in each system and is therefore arbitrary.

In the right hand panel of Fig. \ref{fig:mass_function} we also show the CMF during the starburst at $t\sim 175 \ \rm Myr$, when the young clusters are numerous, binned with 10 clusters per bin. We fit a power-law with best-fit indices of $\alpha=-1.67\pm 0.15$ and $\alpha=-1.96\pm 0.18$ to clusters above masses $10^{2.5}\, M_\odot$ and $10^{3}\, M_\odot$, respectively, to mimic the typical observational sample completeness limits. The fitted power-law indices agree well with the observed slopes in e.g. the Magellanic clouds \citep{2012ApJ...752...96F}. The largest mass-bin includes the most massive globular cluster studied in L19.

\subsection{Cluster formation efficiency}

To further quantify the relationship between star formation and bound cluster formation, we calculate the cluster formation efficiency (CFE or $\Gamma$, \citealt{2008MNRAS.390..759B}) as a function of time for which we use the form
\begin{equation}\label{eq:gamma_sfr}
  \Gamma = \frac{M_\mathrm{*,\, cl}(\tau <10\, \mathrm{Myr})}{M_*(\tau<10 \, \mathrm{Myr})}.
\end{equation}
Here $M_*(\tau<10 \, \mathrm{Myr})$ again includes all stellar mass formed during the last $10$ Myr and $M_\mathrm{*,\, cl}(\tau <10\, \mathrm{Myr})$ includes the stellar mass in clusters with mean stellar ages less than $10$ Myr. We show in Fig. \ref{fig:cfe} the CFE as a function of the $\Sigma_\mathrm{SFR}$ calculated from $100$ pc pixels with SF. The CFE during the simulation is shown in Fig. \ref{fig:cfe} from the first pericentric passage onward, compared to various observed results and analytical predictions. For clarity, we omit from Fig. \ref{fig:cfe} the approach phase and the isolated dwarf as there are only very few clusters forming at any given time, at $10$--$20\%$ efficiency. 

The results in Fig. \ref{fig:cfe} for the simulated CFE follow a similar trend with $\Sigma_\mathrm{SFR}$ as has been predicted analytically for e.g. a typical disk galaxy model in \citet{2012MNRAS.426.3008K}. Observations of the CFE in a number of extra-galactic systems are shown in Fig. \ref{fig:cfe} from a set of references and the results for the Magellanic clouds are highlighted with red symbols. There is some uncertainty in the literature on how the clusters for determining the CFE are selected, which is reflected in e.g. the order of magnitude variance in the observed results for the Magellanic clouds shown in Fig. \ref{fig:cfe}. When older clusters for example with ages between $10$--$100\,\mathrm{Myr}$ are included, the resulting CFE ends up lower due to the environment-dependent cluster destruction which may proceed as steeply as $1/\mathrm{age}$ \citep{2005ApJ...631L.133F} given the presence of disrupting objects such as giant molecular clouds in a gas-rich environment \citep{2006MNRAS.371..793G}. Our results are derived from very young clusters principally to follow the varying SFR, which is also the suggested method in e.g. \citet{2017ApJ...849..128C}. In addition, we focus here on the population of young clusters as we do not detect significant cluster destruction during the duration of our simulation. However, it is unclear whether our stronger than observed cluster survival rate is due to the dwarf galaxy environment or the numerical implementation. 
Finally, one has to bear in mind that we only consider bound objects, whereas in observations young objects may also include open clusters and/or associations that are not bound. 

The CFE during and after the first pericenter agrees with e.g. the results for the Magellanic clouds, which are interacting with the Milky Way and each other, somewhat reminiscent to the early phases of our simulation.
During the merger, the CFE reaches $80$--$90\%$ for the duration of a few Myr, when the most massive clusters are forming and assembling hierarchically. We emphasize that our results for the CFE are absolute upper limits for the formation of bound stellar structures, as our cluster mass is obtained with \subfind\ from 3D data unavailable to observers. The low-mass, high gas-fraction system under study is however an analogue for high-redshift galaxy formation, and high values similar to our CFE in the early Universe have also been reported at least in numerical simulations (e.g. \citealt{2017ApJ...834...69L}).

\subsection{Size and surface density}\label{section:mass-size}

\begin{figure*}
\includegraphics[width=\textwidth]{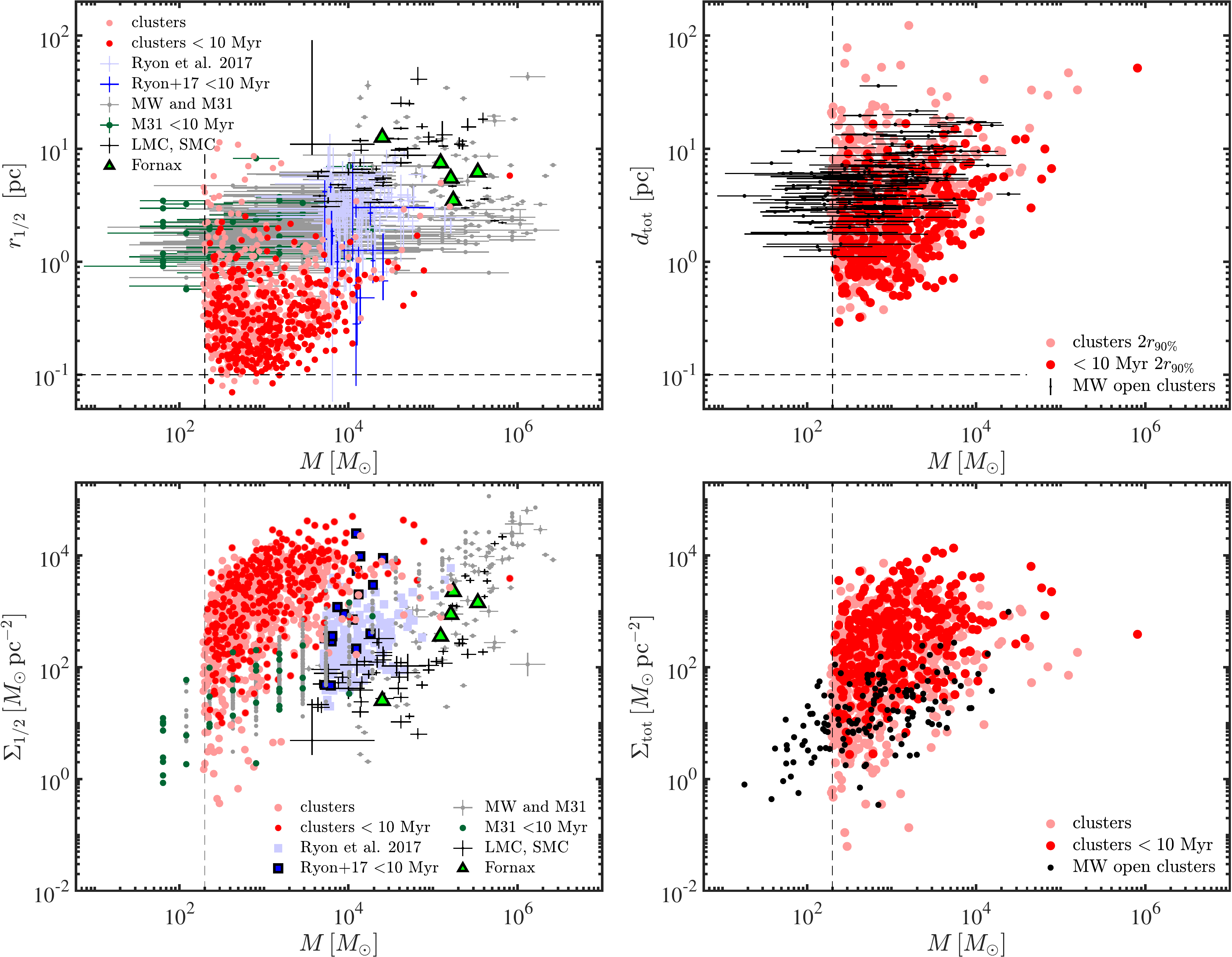}
\caption{Top left: the stellar half mass radius for all (light red) and young (dark red) clusters with respect to stellar mass at simulation time $175$ Myr. The masses and half-light radii for massive clusters in MW, LMC, SMC and the Fornax dwarf spheroidal 
are from \citet{2005ApJS..161..304M}, for M31 in the PHAT survey from \citet{2014ApJ...786..117F} (masses) \citet{2012ApJ...752...95J} (radii), and for clusters in two LEGUS galaxies from \citet{2017ApJ...841...92R}. The minimum cluster mass used in \subfind\ and the softening length of 0.1 pc are indicated with horizontal solid and vertical lines. Top right: diameter measured as two times the radius including $90\%$ of the cluster mass, compared to OCs in the Milky Way from \citet{2006AJ....131.1559V} (based on \citealt{2005ApJ...629..825D}). 
Bottom left: stellar surface density within half-mass (half-light) radius for the data in top left panel. Bottom right: stellar surface density within the clusters for the data in top right panel. \label{fig:size_mass}} 
\end{figure*}

In the left hand panel of Fig. \ref{fig:size_mass} we show the mass and stellar half-mass radius of each of the simulated clusters at a simulation time of $t=175$ Myr, separating the young clusters (darker red symbols) from the older cluster population (lighter red symbols). The simulated clusters are compared to the half-light radii and best-fit masses of clusters in the LMC, SMC, Fornax dwarf spheroidal and the Milky Way, obtained from \citet{2005ApJS..161..304M}, the best-fit effective radii and masses of the M31 clusters from the PHAT survey from \citet{2014ApJ...786..117F} and \citet{2012ApJ...752...95J}, and the effective radii of clusters within two LEGUS galaxies \citep{2017ApJ...841...92R}. In the \citet{2005ApJS..161..304M} data the observed half-light radii have been obtained by fitting an EFF-profile \citep{1987ApJ...323...54E} to the surface brightness profile while the simulated radii are taken directly from the particle data. For the LEGUS galaxies the effective radii are from the best-fit power-law surface brightness parameters obtained with GALFIT \citep{2002AJ....124..266P}, considering only clusters with light profile power-law indices larger than $1.3$. The M31 cluster radii have been obtained by interpolating the flux profiles.

Accurate radii for stellar clusters are often obtained through fitting the surface brightness or star count profile with a radial profile, which is why the left hand panel of Fig. \ref{fig:size_mass} only includes fairly massive clusters. To assess the low mass clusters, the top right panel of Fig. \ref{fig:size_mass} shows the total diameters of our clusters\footnote{Twice the radii including $90\%$ of the stellar mass}, compared to the diameters\footnote{Maximum extent a cluster is separable from the background noise} of open clusters (OCs) in the Milky Way from \citet{2006AJ....131.1559V}, based on data from \citet{2005ApJ...629..825D}.

The sizes of the massive clusters above a few $10^3\,M_\odot$ agree fairly well with the observed $M_*$--$r_{1/2}$ distribution, especially when we look at the young observed population. Similar cluster sizes of $1$--$10$ pc at masses above $10^4\,M_\odot$ have also been found in a high-resolution cluster formation study by \citet{2015MNRAS.446.2038R}. The lower mass clusters seem to be somewhat more compact than their observed counterparts, although the total sizes are in a slightly better agreement with the observed low mass clusters in the top right panel. The small sizes of the low-mass objects might be connected to the employed star formation scheme and will be investigated in greater detail in future work. For some of the smallest clusters the cluster radii are 
also close to the gravitational softening length. For such clusters the internal structure is not fully resolved and thus there is some uncertainty in the cluster sizes, because 
the gravitational motions, although still energy and angular momentum conserving, are softened.

The bottom row of Fig. \ref{fig:size_mass} shows the effective and total mean stellar surface densities from the data points in the top row, either directly from the dataset (WM, LMC, SMC, Fornax with errors) or calculated as $\Sigma_{1/2}=M (2\pi r_{1/2}^2)^{-1}$ and $\Sigma_\mathrm{tot}=M(\pi (d_\mathrm{tot}/2)^2)^{-1}$. Our clusters agree again best with observed young intermediate mass clusters, while the slightly too small radii of the lower-mass clusters translate into higher surface densities than observed. 

There is a tendency for older clusters to have larger sizes and lower mean surface densities, similarly to what is observed in e.g. LMC clusters \citep{1989ApJ...347L..69E, 2003MNRAS.338..120M, 2019arXiv190902049F}. Suggested reasons for the dilution of the central densities include internal stellar dynamics \citep{2019arXiv190902049F}, varying tidal fields \citep{2003MNRAS.343.1025W} and binary black holes \citep{2008MNRAS.386...65M}. Our simulation includes a realistic IMF from $\sim 4 \, M_\odot$ upwards as well as stellar mass black holes in massive SNII remnants, but only resolved down to the gravitational softening length. A more detailed investigation into the dynamical evolution is beyond the scope of this article and will be addressed in future work.

\begin{figure*}
\includegraphics[width=\textwidth]{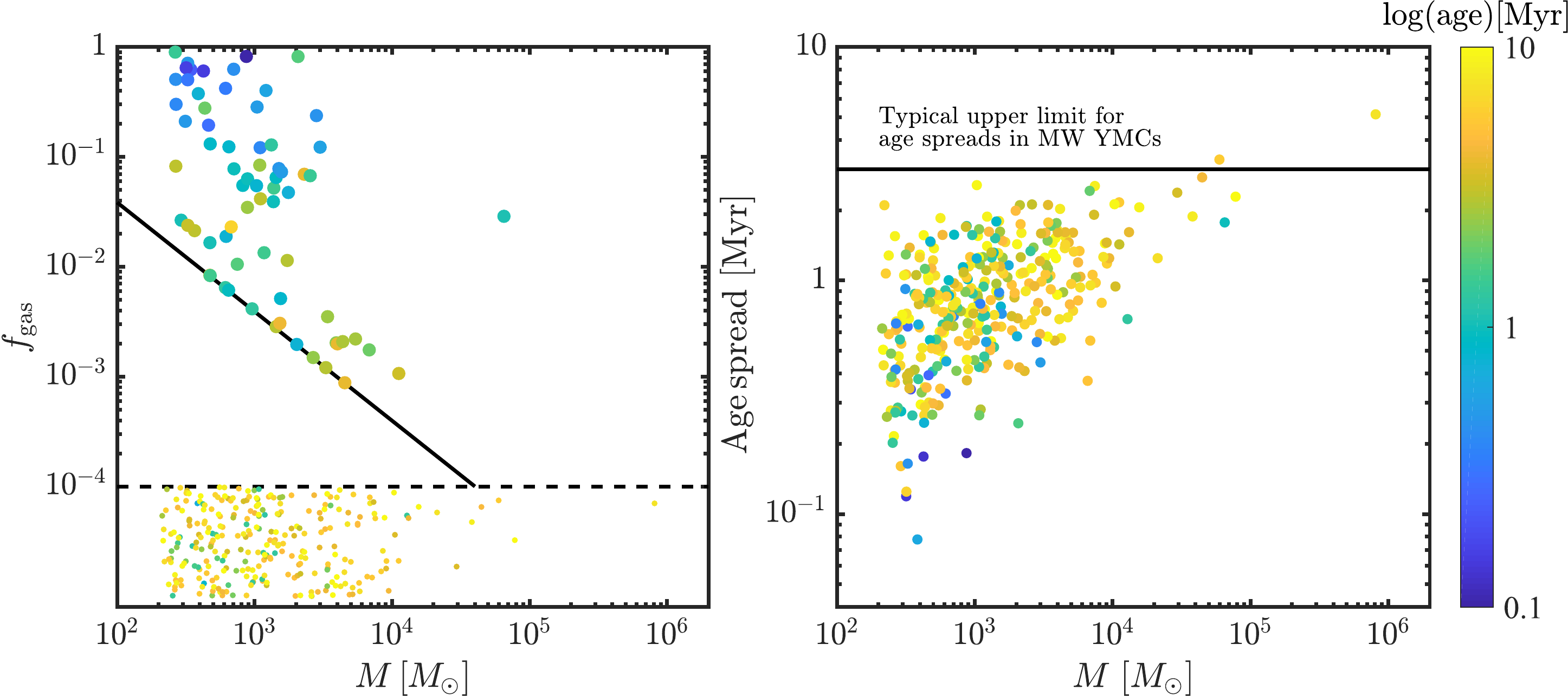}
\caption{Left: Gas fraction, calculated from gas within two half-mass radii from the center of mass of all clusters younger than 10 Myr, at simulation time $175$ Myr. The data points are colored by age. The black solid line shows the limit of one gas particle per cluster, and the randomly scattered points below $f_\mathrm{gas}=10^{-4}$ are clusters with no gas. Right: The spread of ages within $1\sigma$ on either side of the mean stellar age in the young clusters of the left panel. The horizontal line shows a typical upper limit for the age spread in YMCs in the Milky Way \citep{2014prpl.conf..291L}.  \label{fig:gas_fraction}}
\end{figure*}

\subsection{Gas content and age spread}

The star clusters form embedded within filamentary gas clouds, as indicated in Fig. \ref{fig:cluster_overview}. After a few Myr, the stellar feedback in the clusters expel the birth cloud, and the clusters become purely stellar, devoid of gas. Even though only less than $10\%$ of the SNII energy released couples to the ISM \citep{2015MNRAS.451.2757W}, in L19 we found that at least in the massive clusters, the SNII feedback is easily able to deposit more than the binding energy in the gas with typical densities of $\sim 10^3\, M_\odot\,\mathrm{pc}^{-3}$ (see Fig. 3 in L19) when the SNII rates are up to $500$ SNII per Myr per cluster. 

The left panel in Fig. \ref{fig:gas_fraction} shows the mass fraction of gas in each young ($<10$ Myr) cluster at a simulation time of $175$ Myr (as in the right panel of Fig. \ref{fig:mass_function}). The gas fraction has been calculated from gas residing within two stellar half-mass radii of the center of mass of each cluster. Embedded clusters are considered as long as the bound stellar component is more than $\sim200\, M_\odot$ according to \subfind. The gas fractions are shown with respect to the stellar mass, color coded by mean stellar age. Clusters without gas have been spread in the bottom part of the y-axis.

The youngest and least massive clusters are the most gas-rich, and at the age of a few Myr most of the clusters become gas-poor. This result shows how the gas expulsion takes place within the first few Myr of the cluster evolution. As the most extreme example and as also discussed in L19, once the most massive cluster begins to form, it expels all its gas by $7$ Myr into its formation. Stellar feedback before the type II supernovae lowers the total amount of energy required to unbind the gas, which results in gas expulsion fairly soon after the SNII feedback kicks in. Young stellar clusters are very rarely observed within their birth cloud after a few Myr \citep{2003ARA&A..41...57L,2014MNRAS.445..378B, 2015MNRAS.449.1106H,2019MNRAS.490.4648H}, in agreement with the  general trend in Fig. \ref{fig:gas_fraction}.

Assuming a cluster forms from collapsing gas, the duration of the collapse and the time at which a cluster expels its gas set the primary limits for the formation time and age spread of each cluster. When inferring ages of observed clusters, single stellar population models are often used, assuming all stars formed coevally. However, the age distributions in some local stellar clusters have been quantified  on a star by star basis \citep{2014prpl.conf..291L}. Age spreads, when detected, are commonly reported as a few Myr \citep{Bik_2011,Da_Rio_2010}, typically less than $3$ Myr. For massive clusters in external galaxies, the age spread is expected to be no larger than a few 10 of Myrs \citep{2018ARA&A..56...83B}.

We show the $1\sigma$ spread in the ages of the stellar particles within the young simulated clusters in the right panel of Fig. \ref{fig:gas_fraction} as a function of cluster mass and color-coded by cluster age. The spread is here defined as the width of the distribution including $1\sigma$ on either side of the mean stellar age. The clusters seem to form coevally; the age spread in the simulated clusters is typically in the range $\sim 0.1-2$ Myr with only a few clusters with an age spread above 2 Myr, well within the typical values observed in the Milky Way. Even the most massive cluster, which is built
hierarchically, forms its stellar mass on the free-fall timescale, resulting in an age spread of $5.1$ Myr. In other words, the small structures collapse almost simultaneously and merge afterwards. Similar results for stellar age spreads have been reported in other recent numerical studies probing e.g. the high-redshift environment of GC formation such as \citet{2017ApJ...834...69L} and \citet{2019arXiv190611261M}. As in Fig. \ref{fig:gas_fraction}, \citet{2019arXiv190611261M} report higher age spreads with increasing cluster mass, though they study masses above $10^{3.5}\, M_\odot$. 

The derived age spreads are also related to the discussion regarding multiple stellar populations in massive star clusters. Even though we do not include processes necessary to assess the chemical anomalies typically used in the identification of the multiple stellar populations, we can draw some conclusions from the age distribution in the young clusters. Since the clusters become gas-poor at a relatively young age, large amounts of future star formation from enriched material is not supported by our results. Gas accreted from the ISM and expelled by stellar feedback or stellar mass loss could be used for future star formation \citep{2011ApJ...726...36C}, however all of our clusters include little or no gas  ($<10\%$) 
in the 100 Myr old merger remnant. On the other hand, the hierarchical formation of the clusters enables the first forming stars to enrich the infalling material for slightly later star formation during the formation phase. This would require a second enriched population, if present, to be enriched by some fairly rapid processes such as fast-rotating stars or massive binaries \citep{2018ARA&A..56...83B}, rather than e.g. AGB-winds.

\section{Conclusions}\label{section:conclusions}

We have studied the formation of stellar clusters in a high-resolution simulation of a merger between two gas-rich dwarf galaxies, concentrating on the observational properties of the young cluster population. The star formation properties, spanning three orders of magnitude in $\Sigma_\mathrm{SFR}$ during the simulation, were shown to agree with resolved observations of $\Sigma_\mathrm{SFR}$--$\Sigma_\mathrm{gas}$ for galaxies in the present-day universe. The CFR and CFE along the merger correlate with the SFR and $\Sigma_\mathrm{SFR}$, respectively. 

The $\Sigma_\mathrm{SFR}$ during the approach and the first pericentric passage result in similar values to the present-day Magellanic clouds. The young clusters, defined by a mean age of 10 Myr or less, present also comparable values for the correlation between $\Sigma_\mathrm{SFR}$ and the most massive cluster in equivalent age interval in the Magellanic clouds. The CMF of the young SCs, which already has its shape after the first pericentric passage, is similar to a wide range of observed SCs in dwarf galaxies, also outside of the Local Group. The CMF of the young SCs retains its shape all the way through the simulation and the mass of the most massive young cluster increases almost linearly with $\Sigma_\mathrm{SFR}$. We fit a cluster mass function with a power-law index of the order of $\alpha\sim 2$ to the young SC population after the starburst. 

A caveat of our simulations is that we do not detect significant amounts of cluster disruption after their formation, reported in massive galaxies such as the Milky Way and the Antennae (up to $90\%$ per decade, e.g. \citealt{2007AJ....133.1067W}) or the Magellanic clouds \citep{2010ApJ...711.1263C}. Our clusters may form slightly too compact and therefore be more resistant to disruption (see Section \ref{section:mass-size}). This potential caveat will be addressed in future work using modified star formation prescriptions and more detailed $N$-body dynamics. 

Most of the stellar clusters at the massive end of the CMF, as those discussed in L19, have sizes, mean surface densities and stellar age spreads consistent with observed massive young clusters. There is also a slight tendency for older SCs to have larger half-mass radii and lower mean surface densities. The small mass clusters in the simulation seem to form too concentrated, and may therefore be resistant to infant mortality or cluster destruction. In future work we will investigate what is needed to obtain a more realistic size distribution. We will also inspect whether the small sizes are connected to the low cluster disruption rate, which may also result from the dwarf galaxy environment itself. The stellar age spreads in the young SCs are typically between $0.1$ and $2$ Myr. The largest spread of $5.1$ Myr is found in the most massive cluster, which forms hierarchically from tens of smaller clusters along the CMF. 

The merger scenario investigated here resulted in the formation of a stellar cluster population with a mass range from GC-like objects ($8\times 10^5\,M_\odot$) down to a few hundred $M_\odot$. The low shear and shallow gravitational potential of the system consisting of low mass dwarf galaxies enable long life times and easy accretion for even the smaller mass range of SCs. The formation of SCs in interacting dwarf galaxies presents therefore a compelling means for e.g. the origin of ex-situ SCs in outer halos of massive galaxies. 

This study, which is part of the \textsc{griffin} project aimed at resolving the impact of individual massive stars on galaxy evolution, demonstrates that it is now possible to investigate the star cluster populations of entire galaxies, and therefore also the effect of clustered supernova explosions on galaxy evolution, with novel high-resolution numerical simulations. This approach is a challenge for galaxy evolution modelling \citep{2017ARA&A..55...59N} and will hopefully result in novel insights on the underlying processes regulating galaxy evolution. 

\small
\begin{acknowledgements}
N.L. acknowledges the financial support by the Jenny and Antti Wihuri Foundation. T.N. acknowledges support from the DFG cluster of excellence "ORIGINS". N.L. and P.H.J. acknowledge support by the  European Research Council via ERC Consolidator Grant KETJU (no. 818930). C.-Y.H. acknowledges The Center for Computational Astrophysics, supported by the Simons Foundation. S.W. acknowledges support by the European Research Council via ERC Starting Grant RADFEEDBACK (no. 679852) and by the German Science Foundation via CRC956, Project C5. U.P.S. and B.P.M. acknowledge support by an Emmy Noether grant of the Deutsche Forschungsgemeinschaft (DFG, German Research Foundation) under the project number MO 2979/1-1. The computations were carried out at CSC -- IT Center for Science Ltd. in Finland and at Max-Planck Institute for Astrophysics in Germany.

\end{acknowledgements}




\bibliography{references}




\end{document}